\pdfoutput=1
\documentclass[prd,showpacs,letterpaper,10pt,aps,notitlepage,nofootinbib,superscriptaddress]{revtex4-2}

\usepackage{amsfonts,amsmath,graphicx}
\usepackage{hyperref}
\usepackage{color}
\usepackage{placeins}
\usepackage{diagbox}
\usepackage[capitalize]{cleveref}
\usepackage{bm}

\usepackage{xargs}                    % Use more than one optional parameter in a new commands
\usepackage[pdftex,dvipsnames]{xcolor} 
\usepackage[colorinlistoftodos,prependcaption,textsize=small]{todonotes}

\hypersetup{colorlinks, linkcolor = [rgb]{0,0.0,0.75}, citecolor = [rgb]{0,0.0,0.75}, urlcolor = [rgb]{0,0.0,0.75}}

\newcommand{\C}{\mathcal{C}}

\newcommand{\xb}{\textbf{x}}
\newcommand{\yb}{\textbf{y}}
\newcommand{\dagg}{^\dagger}
\newcommand{\op}{\mathcal{O}}
\newcommand{\vp}{\phantom{+}}
\newcommand{\vpz}{\phantom{+0}}
\newcommand{\vz}{\phantom{0}}

\newcommandx{\unsure}[2][1=]{\todo[linecolor=red,backgroundcolor=red!25,bordercolor=red,#1]{#2}}
\newcommandx{\change}[2][1=]{\todo[linecolor=blue,backgroundcolor=blue!25,bordercolor=blue,#1]{#2}}
\newcommandx{\missing}[2][1=]{\todo[linecolor=Orange,backgroundcolor=Orange!25,bordercolor=Orange,#1]{#2}}
\newcommandx{\info}[2][1=]{\todo[linecolor=OliveGreen,backgroundcolor=OliveGreen!25,bordercolor=OliveGreen,#1]{#2}}
\newcommandx{\improve}[2][1=]{\todo[linecolor=Plum,backgroundcolor=Plum!25,bordercolor=Plum,#1]{#2}}

\newcommand{\ltapprox}{\raisebox{-0.5ex}{$\,\stackrel{<}{\scriptstyle\sim}\,$}}

\newcolumntype{C}[1]{>{\centering\arraybackslash}m{#1}}

\definecolor{darkgreen}{rgb}{0,0.5,0}

\begin{document}

\title{Search for \texorpdfstring{$\bm{\bar b \bar b u s}$}{bbar-bbar-u-s} and \texorpdfstring{$\bm{\bar b \bar c u d}$}{bbar-cbar-u-d} tetraquark bound states using lattice QCD}

\author{Stefan Meinel}
\affiliation{Department of Physics, University of Arizona, Tucson, AZ 85721, USA}

\author{Martin Pflaumer}
\affiliation{Goethe-Universit\"at Frankfurt am Main, Institut f\"ur Theoretische Physik, Max-von-Laue-Stra{\ss}e 1, D-60438 Frankfurt am Main, Germany}

\author{Marc Wagner}
\affiliation{Goethe-Universit\"at Frankfurt am Main, Institut f\"ur Theoretische Physik, Max-von-Laue-Stra{\ss}e 1, D-60438 Frankfurt am Main, Germany}
\affiliation{Helmholtz Research Academy Hesse for FAIR, Campus Riedberg, Max-von-Laue-Stra{\ss}e 12, D-60438 Frankfurt am Main, Germany}

\date{\today}

\begin{abstract}
We use lattice QCD to investigate the existence of strong-interaction-stable antiheavy-antiheavy-light-light tetraquarks. We study the $\bar b \bar b u s$ system with quantum numbers $J^P = 1^+$ as well as the $\bar b \bar c u d$ systems with quantum numbers $I(J^P) = 0(0^+)$ and $I(J^P) = 0(1^+)$. We carry out computations on five gauge-link ensembles with $2 + 1$ flavors of domain-wall fermions, including one at the physical pion mass. The bottom quarks are implemented using lattice nonrelativistic QCD, and the charm quarks using an anisotropic clover action. In addition to local diquark-antidiquark and local meson-meson interpolating operators, we include nonlocal meson-meson operators at the sink, which facilitates the reliable determination of the low-lying energy levels. We find clear evidence for the existence of a strong-interaction-stable $\bar b \bar b u s$ tetraquark with binding energy $(-86 \pm 22 \pm 10) \, \text{MeV}$ and mass $(10609 \pm 22 \pm 10) \, \text{MeV}$. For the $\bar b \bar c u d$ systems we do not find any indication for the existence of bound states, but cannot rule out their existence either.
\end{abstract}

\maketitle

% ********************
% ********************
% ********************
% ********************
% ********************

\section{Introduction}

Hadrons with integer spin, in particular those corresponding to low-lying states in the respective spectra, are typically ordinary mesons composed of a single valence quark and a single valence antiquark. They might, however, also contain two valence quarks and two valence antiquarks. Such so-called tetraquarks\footnote{In the literature, the term ``tetraquark'' is somewhat ambiguous. In certain papers it exclusively refers to a diquark-antidiquark structure, while in other papers it is used more generally for arbitrary bound states and resonances with a strong four-quark component, including e.g.\ mesonic molecules. Throughout this paper we follow the latter convention.} were discovered only recently, primarily in the heavy-quark sector \cite{Belle:2011aa,Olsen:2015zcy,Lebed:2016hpi,Esposito:2016noz,Richard:2016eis,Olsen:2017bmm,Brambilla:2019esw,Chen:2022asf}. Of particular importance is the recent discovery of an anticharm-anticharm-light-light tetraquark $T_{cc}$ by the LHCb collaboration with isospin $I = 0$ and mass slightly below the lowest two-meson threshold corresponding to $D D^\ast$ \cite{LHCb:2021vvq,LHCb:2021auc}. Such antiheavy-antiheavy-light-light systems $\bar Q \bar Q q q$ are \emph{manifestly} flavor-exotic and are simpler to investigate theoretically than their $\bar Q Q \bar q q$ counterparts, because the lowest relevant decay threshold consists of a pair of heavy-light mesons, typically with similar mass, and not the significantly lighter scattering states containing a light meson and ordinary quarkonium (or even the annihilation products of the quarkonium). Moreover, strong-interaction-stable $\bar{Q} \bar{Q} q q$ tetraquarks are expected to exist for sufficiently large heavy quark masses $m_Q$ \cite{Carlson:1987hh,Manohar:1992nd,Eichten:2017ffp}. In this limit, the two heavy antiquarks form a color-triplet with size of order $(\alpha_s m_Q)^{-1}$ and binding energy of order $\alpha_s^2 m_Q$ due to the attractive Coulomb potential at small $\bar{Q} \bar{Q}$ separations. $\bar{Q} \bar{Q} q q$ tetraquarks are then quite similar to heavy-light-light baryons $Q q q$, just like heavy-heavy-light baryons $\bar{Q} \bar{Q} \bar{q}$ are related to heavy-light mesons $Q \bar{q}$ \cite{Savage:1990di,Brambilla:2005yk,Cohen:2006jg,Mehen:2017nrh}. Thus, the question is whether the physical heavy quark mass $m_c$ or $m_b$ is sufficiently large for $\bar{Q} \bar{Q} q q$ bound states to exist below the corresponding lowest $\bar{Q} q$-$\bar{Q} q$ two-meson thresholds.

Following initial studies using potential models, effective field theories, and QCD sum rules \cite{Carlson:1987hh, Manohar:1992nd, SilvestreBrac:1993ss, Brink:1998as, Vijande:2003ki, Janc:2004qn, Vijande:2006jf, Navarra:2007yw, Ebert:2007rn, Zhang:2007mu, Lee:2009rt, Karliner:2017qjm, Eichten:2017ffp, Wang:2017uld, Richard:2018yrm, Park:2018wjk, Wang:2018atz, Liu:2019stu}, as well as analyses based on static meson-meson potentials from lattice QCD \cite{Bicudo:2012qt,Brown:2012tm,Bicudo:2015kna,Bicudo:2015vta,Bicudo:2016ooe,Bicudo:2017szl}, direct lattice-QCD calculations with finite-mass $b$-quarks implemented using lattice NRQCD have now firmly established the existence of a stable $\bar{b} \bar{b} u d$ tetraquark with quantum numbers $I(J^P) = 0(1^+)$  \cite{Francis:2016hui,Junnarkar:2018twb,Leskovec:2019ioa,Mohanta:2020eed}. $\bar{Q} \bar{Q} q q$ systems with different flavor combinations have also been explored. Lattice calculations by two independent groups \cite{Francis:2016hui,Junnarkar:2018twb} yield agreement that there is a strong-interaction-stable $\bar{b} \bar{b} u s$ tetraquark with $J^P = 1^+$ and binding energy around $-80 \, \text{MeV} \ldots -100 \, \text{MeV}$. There is more variation among non-lattice approaches, with Refs.\ \cite{SilvestreBrac:1993ss,Lee:2009rt,Eichten:2017ffp,Wang:2017uld,Park:2018wjk,Deng:2018kly,Braaten:2020nwp,Dai:2022ulk} predicting a stable $\bar{b} \bar{b} u s$ tetraquark while Refs.\ \cite{Ebert:2007rn,Lu:2020rog,Faustov:2021hjs} conclude the opposite, in contradiction with the aforementioned lattice-QCD results. Another interesting four-quark system is the $\bar{b} \bar{c} u d$ system with $I(J^P) = 0(1^+)$, which was also investigated using lattice QCD. In this more challenging case, independent groups have so far arrived at different conclusions. In Ref.\ \cite{Francis:2018jyb} the existence of a strong-interaction-stable tetraquark was reported, but later revoked \cite{Hudspith:2020tdf}, while other authors find indication for its existence \cite{Padmanath:2021qje}. Also other approaches do not exhibit a consistent picture. Refs.\ \cite{Lee:2009rt,Chen:2013aba,Karliner:2017qjm,Sakai:2017avl,Agaev:2018khe,Deng:2018kly,Carames:2018tpe,Yang:2019itm,Tan:2020ldi} predict the existence of such a tetraquark, while Refs.\ \cite{Ebert:2007rn,Eichten:2017ffp,Park:2018wjk,Braaten:2020nwp,Lu:2020rog} claim the opposite. Clearly, further precision lattice QCD studies of this system are highly desirable.

Note that $\bar Q \bar Q q q $ tetraquarks with heavy $\bar b$ quarks have not yet been observed experimentally. However, possible search strategies are discussed in Refs.\ \cite{Moinester:1995fk,Ali:2018ifm,Ali:2018xfq}. As mentioned above, the closely related $T_{cc}$ tetraquark with quark content $\bar{c} \bar{c} u d$ was recently discovered by the LHCb collaboration \cite{LHCb:2021vvq,LHCb:2021auc}. A first lattice-QCD study of this system at a heavier-than-physical pion mass can be found in Ref.\ \cite{Padmanath:2022cvl}. 

In this work we focus on the $\bar b \bar b u s$ system with quantum numbers $J^P = 1^+$ and the $\bar b \bar c u d$ systems with quantum numbers $I(J^P) = 0(0^+)$ and $I(J^P) = 0(1^+)$. We employ the same lattice QCD setup as in our previous study of the $\bar b \bar b u d$ tetraquark with quantum numbers $I(J^P) = 0(1^+)$ \cite{Leskovec:2019ioa}, i.e.\ we use NRQCD to discretize $\bar{b}$ quarks and domain-wall light quarks. The charm quarks, which were not part of our previous study, are implemented using an anisotropic clover action with three parameters tuned nonperturbatively to eliminate heavy-quark discretization errors. In the construction of the two-point correlation functions, we consider not only local interpolating operators (in which the four quarks are jointly projected to zero momentum, i.e.\ where each quark is centered around the same point in space), but also non-local interpolating operators (in which each of the two quark-antiquark pairs forming a color-singlet is projected to zero momentum individually). It has been shown in previous studies of other four-quark systems that including both types of interpolating operators is required to reliably determine ground state energies in exotic channels \cite{Mohler:2013rwa,Lang:2014yfa,Leskovec:2019ioa}. In this way we expand on the works of Refs.\ \cite{Francis:2016hui,Francis:2018jyb,Junnarkar:2018twb,Hudspith:2020tdf,Padmanath:2021qje}, where non-local interpolating operators were not considered.

This article is organized in the following way. In \cref{sec:latticesetup} we briefly summarize our lattice setup. In \cref{SEC597} we discuss the interpolating operators for the three systems we investigate and the corresponding correlation functions. In \cref{SEC332} we give the lattice results for the single heavy-light meson energies. \cref{sec:results} is the main section, where we present our numerical results for the antiheavy-antiheavy-light-light four-quark systems. We explore the importance of each of our interpolating operators, extract finite-volume energy levels for all ensembles, and formulate conclusions concerning the existence of 
antiheavy-antiheavy-light-light tetraquarks at the physical $u$ and $d$ quark mass and in infinite spatial volume. We summarize the main points of our work in \cref{SEC596} and give a brief outlook. Note that results obtained at an early stage of this project were presented at recent conferences \cite{Pflaumer:2020ogv,Pflaumer:2021ong}.

% ********************
% ********************
% ********************
% ********************
% ********************

\section{\label{sec:latticesetup}Lattice setup}

% ********************
% ********************
% ********************

\subsection{Gauge link configurations, light quark and bottom quark propagators}

The computations presented in this work were carried out on five ensembles of gauge link configurations generated by the RBC and UKQCD collaborations \cite{Aoki:2010dy, Blum:2014tka} using the Iwasaki gauge action \cite{Iwasaki:1984cj} and $N_f=2+1$ flavors of domain-wall fermions \cite{Kaplan:1992bt, Shamir:1993zy, Furman:1994ky, Brower:2012vk}. The ensembles differ in the lattice spacing, the lattice size and the pion mass and are summarized in \cref{tab:configurations}. Further details can be found in our previous lattice QCD study of a $\bar b \bar b u d$ tetraquark \cite{Leskovec:2019ioa}, where we used exactly the same ensembles.

\begin{table}[htb]
	\centering
	\begin{tabular}{cccccccc} \hline \hline 
		Ensemble & $N_s^3 \times N_t$ & $a$ [fm] 	& $a m_{u;d}^{(\rm sea;val)}$ & $a m_{s}^{(\rm sea)}$  & $a m_{s}^{(\rm val)}$ 	& $m_\pi$ [MeV]  & $N_{\textrm{samples}}$ \\ \hline
		C00078 & $48^3 \times 96$& $0.1141(3)$ & $0.00078$           & $0.0362$ & $0.0362$	& $139(1)$ & $\phantom{0}2560$ sl, $\phantom{0}80$  ex \\ 
		C005 & $24^3 \times 64$	 & $0.1106(3)$ & $0.005\phantom{00}$ & $0.04\phantom{00}$ & $0.0323$ & $340(1)$ & $\phantom{0}9952$ sl, $311$ ex \\
		C01	 & $24^3 \times 64$	 & $0.1106(3)$ & $0.01\phantom{000}$ & $0.04\phantom{00}$ & $0.0323$ & $431(1)$ & $\phantom{0}9056$ sl, $283$ ex \\ 
		F004 & $32^3 \times 64$  & $0.0828(3)$ & $0.004\phantom{00}$ & $0.03\phantom{00}$ & $0.0248$ & $303(1)$ & $\phantom{0}8032$ sl, $251$ ex \\
		F006 & $32^3 \times 64$  & $0.0828(3)$ & $0.006\phantom{00}$ & $0.03\phantom{00}$ & $0.0248$ & $360(1)$ & $14144$ sl, $442$ ex \\ \hline \hline& 
	\end{tabular}
	\caption{\label{tab:configurations}Gauge link ensembles \cite{Aoki:2010dy, Blum:2014tka} and light quark propagators used in this work. $N_s$, $N_t$: number of lattice sites in spatial and temporal directions; $a$: lattice spacing; $a m_{q}^{(\rm sea)}$: sea-quark mass of flavor $q$; $a m_{q}^{(\rm val)}$: valence-quark mass of flavor $q$; $m_\pi$: pion mass. We use all-mode-averaging
\cite{Blum:2012uh,Shintani:2014vja} with 32 or 64 sloppy (sl) and 1 or 2 exact (ex) samples per configuration, leading to the total numbers of samples given in the last column of the table.}
\end{table}

We use point-to-all propagators with Gaussian-smeared sources (cf.~Sec.~\ref{SEC488}). We employ the all-mode averaging technique \cite{Blum:2012uh,Shintani:2014vja} with 32 or 64 sloppy (sl) and 1 or 2 exact (ex) samples per configuration, where the sloppy correlation-function samples differ from the exact samples in that they use light and strange propagators computed with a reduced solver iteration count. The light-quark propagators are identical to those used in Ref.~\cite{Leskovec:2019ioa}. The valence strange-quark masses are close to the physical value \cite{Blum:2014tka}. For the bottom quarks we use lattice NRQCD \cite{ThackerLepage91,Lepage:1992tx}; also here the setup is the same as in Ref.\ \cite{Leskovec:2019ioa}.

% ********************
% ********************
% ********************

\subsection{Charm quark propagators}

For the charm quarks we use an anisotropic clover action, following the approach developed in Refs.~\cite{El-Khadra:1996wdx,Chen:2000ej,Aoki:2001ra,Aoki:2003dg,Christ:2006us,Lin:2006ur,RBC:2012pds}, which allows the removal of discretization errors of order $|a\mathbf{p}|$, $(am)^n$, and $|a\mathbf{p}|(am)^n$ for all non-negative integers $n$. Specifically, our action is of the same form as in Ref.~\cite{RBC:2012pds}, and we tuned the mass $a m_c$ (denoted as $a m_0$ in Ref.~\cite{RBC:2012pds}), anisotropy parameter $\zeta$, and clover coefficient $c_P$ nonperturbatively such that the $D_s$ meson rest mass, kinetic mass, and hyperfine splitting extracted from two-point functions on each ensemble match the experimental values \cite{ParticleDataGroup:2020ssz}. These observables calculated on each ensemble are found to agree with experiment within 0.4\%, 1.0\%, and 1.4\% (or better) precision, respectively. The values of the action parameters are given in Table \ref{tab:charmparams}.

\begin{table}[htb]
  \begin{tabular}{lcccccc}
    \hline\hline
    Ensemble & \hspace{2ex} & $ a m_c$ & \hspace{2ex} & $\zeta$ & \hspace{2ex} & $c_P$ \\
    \hline
    C00078                   && $0.2751$  && $1.1883$ && $2.0712$ \\
    C005, C01                && $0.1541$  && $1.2004$ && $1.8407$ \\
    F004, F006               && $-0.0517$ && $1.1021$ && $1.4483$ \\
    \hline\hline
  \end{tabular}
  \caption{\label{tab:charmparams}Parameters used in the anisotropic clover action for the charm quarks. The form of the heavy-quark action is given in Ref.~\cite{RBC:2012pds}, where $m_c$ is denoted by $m_0$.}
\end{table}

% ********************
% ********************
% ********************
% ********************
% ********************

\section{\label{SEC597}Interpolating operators and correlation functions}

% ********************
% ********************
% ********************

\subsection{Four-quark systems \label{sec:fourQuarkSystems}}

The main goal of this work is to compute low-lying energy levels of antiheavy-antiheavy-light-light four-quark systems with quark content $ \bar{b}\bar{b}us $ and $ \bar{b}\bar{c}ud $ and to explore whether the ground-state energies are below the lowest corresponding meson-meson thresholds. A ground-state energy sufficiently far below threshold (compared to the expected size of finite-volume effects) would indicate a four-quark bound state, i.e.\ the existence of a strong-interaction-stable tetraquark. In the $ \bar{b}\bar{b}us $ case we consider exclusively the $ J^P=1^+ $ channel, which is the only channel where one can expect sufficiently strong attractive forces to generate a bound state (see the symmetry arguments given in Section~III.B of Ref.\ \cite{Bicudo:2016ooe}). In the $ \bar{b}\bar{c}ud $ case we focus on $I = 0$, again because of the related stronger attraction of the four quarks \cite{Bicudo:2015kna,Eichten:2017ffp}. There are two promising $I = 0$ channels, because the heavy antiquark pair $ \bar{b}\bar{c} $ can be either flavor symmetric or flavor antisymmetric. The symmetric $ I(J^P)=0(1^+) $ channel is conceptually similar to the $ J^P=1^+ $ channel for $ \bar{b}\bar{b}us $ (and also for $ \bar{b}\bar{b}ud $, as investigated in detail within the same setup in our previous work \cite{Leskovec:2019ioa}), while the antisymmetric $ I(J^P)=0(0^+) $ channel is different.

To be able to resolve possibly existing four-quark bound states as well as meson-meson scattering states, we employ both local interpolating operators and non-local (``scattering'') interpolating operators. Local operators are constructed from products of four quark fields at the same point in space, followed by projection of the product to total momentum zero. Scattering operators, on the other hand, resemble two heavy-light mesons with independent spatial locations and individual projection of each meson to momentum zero. Local interpolating operators can be categorized further into meson-meson and diquark-antidiquark operators. The local meson-meson operators (as well as the scattering operators) resemble pairs of mesons with overall quantum numbers identical to those of the four-quark system of interest. For each local meson-meson operator we also consider a corresponding scattering operator, which differs only in the momentum projection. The importance of diquark-antidiquark pairs was pointed out in Refs.\ \cite{Jaffe:2004ph,Cheung:2017tnt,Bicudo:2021qxj}. Following Jaffe's notation of ``good'' and ``bad'' diquarks \cite{Jaffe:2004ph}, our diquark-antidiquark operators are designed in such a way that the light diquark ($ us $ or $ ud $) is a ``good'' diquark. If possible, we choose for the heavy diquark also a ``good'' configuration (in the case of $\bar{b}\bar{c}ud$ with $ I(J^P)= 0(0^+) $), otherwise we use a ``bad'' heavy diquark (for $\bar{b}\bar{b}us$ and for $\bar{b}\bar{c}ud$ with $ I(J^P)= 0(1^+) $).

As we demonstrated in our previous work \cite{Leskovec:2019ioa}, scattering operators play an important role in extracting low-lying energy levels, because they generate sizable overlaps to energy eigenstates close to two-meson thresholds. In particular, if a four-quark bound state exists, scattering operators can eliminate contamination in the fit result for the corresponding energy level caused by nearby scattering states.

% ********************

\subsubsection{Interpolating operators for $ \bar{b}\bar{b}us $ with $ J^P=1^+ $ \label{sec:fourQuarkSystems_bbus}}

% B  5279 / 5280
% B* 5325  (d = 45)
% Bs  5367
% Bs* 5415  (d = 49)
% |(Bs* - Bs) - (B* - B)| < 6 MeV

In contrast to the $ \bar{b}\bar{b}ud $ system discussed in Ref.\ \cite{Leskovec:2019ioa}, which has an $ \text{SU}(2) $ isospin symmetry, there is no such symmetry for the light $u s$ quarks for the $ \bar{b}\bar{b}us $ system. The consequence is that there are not only two, but three relevant meson-meson thresholds, which are rather close, within around $50 \, \text{MeV}$. They correspond to $ B B_s^\ast $, $ B^\ast B_s $ (which is around $5 \, \text{MeV}$ above $ B B_s^\ast $) and  $ B^\ast B_s^\ast $ (which is around $50 \, \text{MeV}$ above $ B B_s^\ast $). The corresponding four local interpolating operators (three meson-meson operators and one diquark-antidiquark operator) are
\begin{align}
\label{eq:op_BBsast_total_zero} &\op_1 = \op_{[B B_s^\ast](0)} =\sum_{\xb}	\bar{b}\gamma_5 u(\xb) \, \bar{b}\gamma_j s(\xb)	 \\
\label{eq:op_BastBs_total_zero} &\op_2 = \op_{[B^\ast B_s](0)} =\sum_{\xb}	\bar{b}\gamma_j u(\xb) \, \bar{b}\gamma_5 s(\xb)	 \\
\label{eq:op_BastBsast_total_zero} &\op_3 = \op_{[B^\ast B_s^\ast](0)} =  \epsilon_{j k l} \sum_{\xb}	\bar{b}\gamma_k u(\xb) \, \bar{b}\gamma_l s(\xb)	 \\
\label{eq:op_Dd_total_zero}&\op_4 = \op_{[D d](0)} =\sum_{\xb}\bar{b}^a \gamma_j \C \bar{b}^{b,T}(\xb)\, u^{a,T}  \C \gamma_5  s^b(\xb) ,
\end{align}
and the three scattering operators are
\begin{align}
\label{eq:op_BBsast_sep_zero} &\op_5 = \op_{B(0) B_s^\ast(0)} = \bigg(\sum_{\xb}	\bar{b}\gamma_5 u(\xb)\bigg) \, \bigg(\sum_{\yb}\bar{b}\gamma_j s(\yb)\bigg) \\
\label{eq:op_BastBs_sep_zero} &\op_6 = \op_{B^\ast(0) B_s(0)} = \bigg(\sum_{\xb}	\bar{b}\gamma_j u(\xb)\bigg) \, \bigg(\sum_{\yb} \bar{b}\gamma_5 s(\yb)\bigg) \\
\label{eq:op_BastBsast_sep_zero} &\op_7 = \op_{B^\ast(0) B_s^\ast(0)} = \epsilon_{j k l} \bigg(\sum_{\xb}	\bar{b}\gamma_k u(\xb)\bigg) \, \bigg(\sum_{\yb} \bar{b}\gamma_l s(\yb)\bigg) .
\end{align}
Above, $ a,b $ are color indices, $ j,k,l $ are spatial indices, and $ \C= \gamma_0 \gamma_2 $ is the charge conjugation matrix.

We note that the operators $ \op_3 $, $ \op_4 $ and $ \op_7 $ are antisymmetric in the light quark flavors. The operators $ \op_1 $ and $ \op_2 $ as well as the operators $ \op_5 $ and $ \op_6 $ can be linearly combined in such a way that there is one symmetric and one antisymmetric light flavor combination.

% ********************

\subsubsection{\label{sec:ops_bcudJ0}Interpolating operators for $ \bar{b}\bar{c}ud $ with $ I(J^P)= 0(0^+) $}

% D  1865 / 1870
% D* 2007 / 2010  (d = 142 / 145)
% B  5279 / 5280
% B* 5325  (d = 45)

The lowest meson-meson thresholds in this channel are $ B D $ and $ B^\ast D^\ast $. Their energy difference is, however, sizable, approximately $ 190 \, \text{MeV} $. Thus, we expect that resolving energy levels close to the $ B^\ast D^\ast $ threshold is not of central importance when studying this channel and exploring the possible existence of a four-quark bound state below the $ B D $ threshold. Consequently, we only consider a single meson-meson structure of $B D$ type. The corresponding two local operators are
\begin{align}
\label{eq:op_BD_total_zero} &\op_1 = \op_{[B D](0)} =\sum_{\xb}	\bar{b}\gamma_5 u(\xb) \, \bar{c}\gamma_5 d(\xb) - (d \leftrightarrow u) 	 \\
&\op_2 = \op_{[D d](0)} =\sum_{\xb}\bar{b}^a \gamma_5 \C \bar{c}^{b,T}(\xb)\, u^{a,T}  \C \gamma_5  d^b(\xb) - (d \leftrightarrow u) ,
\end{align}
and the only scattering operator is
\begin{align}
\label{eq:op_BD_sep_zero} &\op_3 = \op_{B(0)D(0)} =\bigg(\sum_{\xb}	\bar{b}\gamma_5 u(\xb)\bigg) \, \bigg(\sum_{\yb}\bar{c}\gamma_5 d(\yb)\bigg)	- (d \leftrightarrow u) .
\end{align}
The quantum number $I = 0$ implies the antisymmetric light flavor combination $u d - d u$ (as in our previous study \cite{Leskovec:2019ioa} of the $ \bar{b} \bar{b} u d $ system). The heavy quark flavors $ \bar{b} \bar{c} $ are also in an antisymmetric combination, allowing $ J=0 $, which is not possible for heavy quark flavors $ \bar{b} \bar{b} $.

% ********************

\subsubsection{\label{sec:ops_bcudJ1}Interpolating operators for $ \bar{b}\bar{c}ud $ with $ I(J^P)=0(1^+) $}

For total angular momentum $ J=1 $ the lowest meson-meson thresholds are $ B^\ast D $, $ B D^\ast $ and $ B^\ast D^\ast $. We follow a similar strategy as in the previous subsection and do not consider a $ B^\ast D^\ast $ meson-meson structure. The other two thresholds are separated by approximately $100\,\textrm{MeV} $. Thus, we use the three local operators
\begin{align}
\label{eq:op_BastD_total_zero} &\op_1 = \op_{[B^\ast D](0)} =\sum_{\xb}	\bar{b}\gamma_j u(\xb) \, \bar{c}\gamma_5 d(\xb) - (d \leftrightarrow u) 	 \\
\label{eq:op_BDast_total_zero} &\op_2 = \op_{[B D^\ast](0)} =\sum_{\xb}	\bar{b}\gamma_5 u(\xb) \, \bar{c}\gamma_j d(\xb) - (d \leftrightarrow u) 	 \\
&\op_3 = \op_{[D d](0)} =\sum_{\xb}\bar{b}^a \gamma_j \C \bar{c}^{b,T}(\xb)\, u^{a,T}  \C \gamma_5  d^b(\xb) - (d \leftrightarrow u) ,
\end{align}
and the two scattering operators
\begin{align}
\label{eq:op_BastD_sep_zero} &\op_4 = \op_{B^\ast(0)D(0)} =\bigg(\sum_{\xb}	\bar{b}\gamma_j u(\xb)\bigg) \, \bigg(\sum_{\yb}\bar{c}\gamma_5 d(\yb)\bigg)	- (d \leftrightarrow u)\\
\label{eq:op_BDast_sep_zero} &\op_5 = \op_{B(0)D^\ast(0)} =\bigg(\sum_{\xb}	\bar{b}\gamma_5 u(\xb)\bigg) \, \bigg(\sum_{\yb}\bar{c}\gamma_j d(\yb)\bigg)	- (d \leftrightarrow u) .
\end{align}

% ********************

\subsubsection{\label{SEC488}Quark propagators and correlation functions}

As in our previous work \cite{Leskovec:2019ioa} we apply standard smearing techniques to improve the overlap generated by the interpolating operators to the low-lying energy eigenstates. All quark-fields in \cref{eq:op_BBsast_total_zero} to \cref{eq:op_BDast_sep_zero} are Gaussian-smeared,
\begin{equation}
q_{\textrm{smeared}}=\left(1 + \frac{\sigma^2_{\textrm{Gauss}}}{4N_{\textrm{Gauss}}}\Delta\right)^{N_{\textrm{Gauss}}} q ,
\end{equation}
where $ \Delta $ is the nearest-neighbor gauge-covariant spatial Laplacian. For the Gaussian smearing of the up, down, and strange quarks we use APE-smeared spatial gauge links \cite{Albanese:1987ds}\footnote{A single sweep of APE smearing with parameter $\alpha_\textrm{APE}$ is defined as in Eq.~(8) of Ref.~\cite{Bonnet:2000dc}, and we apply $N_\textrm{APE}$ such sweeps.}, while for the charm quarks we use stout-smeared spatial gauge links \cite{Morningstar:2003gk}. The reason for using different types of smearing is that we reuse quark propagators computed previously for other projects. No link smearing is used in the bottom quarks. All smearing parameters are listed in \cref{tab:smearingParams}.

\begin{table}[htb]
	\centering
	\begin{tabular}{lccccccccccccccccc} \hline \hline
		Ensemble & \multicolumn{4}{c}{Up and down quarks}&& \multicolumn{4}{c}{Strange quarks} && \multicolumn{4}{c}{Charm quarks} && \multicolumn{2}{c}{Bottom quarks}\\ 
		& $N_\textrm{Gauss}$ & $\sigma_\textrm{Gauss}$ & $N_\textrm{APE}$ & $\alpha_\textrm{APE}$ &\hspace{2ex}	% up,down
		& $N_\textrm{Gauss}$ & $\sigma_\textrm{Gauss}$ & $N_\textrm{APE}$ & $\alpha_\textrm{APE}$ &	\hspace{2ex}										% strange
		& $N_\textrm{Gauss}$ & $\sigma_\textrm{Gauss}$ & $N_\textrm{stout}$ & $\rho$&	\hspace{2ex}			% charm
		& $N_\textrm{Gauss}$ & $\sigma_\textrm{Gauss}$  \\ \hline									% bottom
		C00078 	   & $ 100 $         & $7.171$ & $25$ & $2.5$ && $ 30 $ & $ 4.350 $ & $25$ & $2.5$ && $ 10 $ & $ 2.00 $ &  $ 10 $ & $ 0.08 $ && $ 10 $ & $ 2.0 $ \\
		C005, C01  & $\phantom{0}30$ & $4.350$ & $25$ & $2.5$ && $ 30 $ & $ 4.350 $ & $25$ & $2.5$ && $ 10 $ & $ 2.00 $ &  $ 10 $ & $ 0.08 $ && $ 10 $ & $ 2.0 $ \\
		F004, F006 & $\phantom{0}60$ & $5.728$ & $25$ & $2.5$ && $ 60 $ & $ 5.728 $ & $25$ & $2.5$ && $ 16 $ & $ 2.66 $ &  $ 10 $ & $ 0.08 $ && $ 10 $ & $ 2.0 $ \\ \hline \hline
	\end{tabular}
\caption{\label{tab:smearingParams} Parameters for the smearing of quark-fields.}
\end{table}

For each of the three systems discussed in \cref{sec:fourQuarkSystems_bbus} to \cref{sec:ops_bcudJ1} we computed temporal correlation matrices 
\begin{equation}
C_{jk} (t) = \Big\langle \op_j(t) \op_k\dagg(0) \Big\rangle
\label{eq:defCorrelationMatrix}
\end{equation}
($ \langle\dots \rangle $ denotes the expectation value of the lattice QCD path integral, and $j$, $k$ now label different operator structures), from which we determine the low-lying energy eigenvalues and obtain information about the quark composition of the corresponding eigenstates, as discussed in detail in \cref{sec:results}.

All computations are based on point-to-all propagators with sources smeared as discussed above. For the light quarks we used the same propagators as in our previous work \cite{Leskovec:2019ioa}, where further technical details are discussed. As a consequence, we are restricted to correlation functions with a local interpolating operators at the source, for which one can use translational invariance to replace the spatial sum by a simple multiplication with the spatial volume. At the sink, however, both local and non-local interpolating operators are used. Thus, our correlation matrices are non-square matrices of sizes $7 \times 4$, $3 \times 2$ and $5 \times 3$, respectively, for the systems discussed in \cref{sec:fourQuarkSystems_bbus} to \cref{sec:ops_bcudJ1}. It is straightforward to show that all three correlation matrices are real-valued and that the square sub-matrices are symmetric. We verified that our numerical results are consistent with these properties and exploited them to increase statistical precision. Similarly, we used the time reversal symmetry to relate $ C_{jk}(t) $ and $ C_{jk}(-t) $, which reduces statistical uncertainties even further.

% ********************

\subsection{$ B $, $ B_s $ and $ D $ mesons}

In \cref{sec:results} we will compare the resulting ground state energies of the $ \bar{b}\bar{b}us $ and $ \bar{b}\bar{c}ud $ four-quark systems discussed above to the respective lowest meson-meson thresholds. To this end, we also computed the energies of the pseudoscalar and vector $ B $, $ B_s $, and $ D $ mesons using exactly the same setup. The corresponding interpolating operators are
\begin{alignat}{2}
\label{eq:Binterpolator}  &\op_{B(0)}				&&= \sum_{\xb} \bar{b}(\xb) \gamma_5 u(\xb) , \\
\label{eq:BastInterpolator} & \op_{B^\ast(0)} 		&&= \sum_{\xb} \bar{b}(\xb) \gamma_j u(\xb) , \\
\label{eq:Bsinterpolator} & \op_{B_s(0)} 			&&= \sum_{\xb} \bar{b}(\xb) \gamma_5 s(\xb) , \\
\label{eq:BsastInterpolator} & \op_{B_s^\ast(0)} 	&&= \sum_{\xb} \bar{b}(\xb) \gamma_j s(\xb) , \\
\label{eq:Dinterpolator} & \op_{D(0)} 				&&= \sum_{\xb} \bar{c}(\xb) \gamma_5 u(\xb) , \\
\label{eq:Dastinterpolator} & \op_{D^\ast(0)} 			&&= \sum_{\xb} \bar{c}(\xb) \gamma_j u(\xb) .
\end{alignat}

% ********************
% ********************
% ********************
% ********************
% ********************

\section{\label{SEC332}Energies of pseudoscalar and vector $ B $, $ B_s $ and $ D $ mesons}

\begin{figure}[htb]
	\centering
	\includegraphics[width=0.45\linewidth]{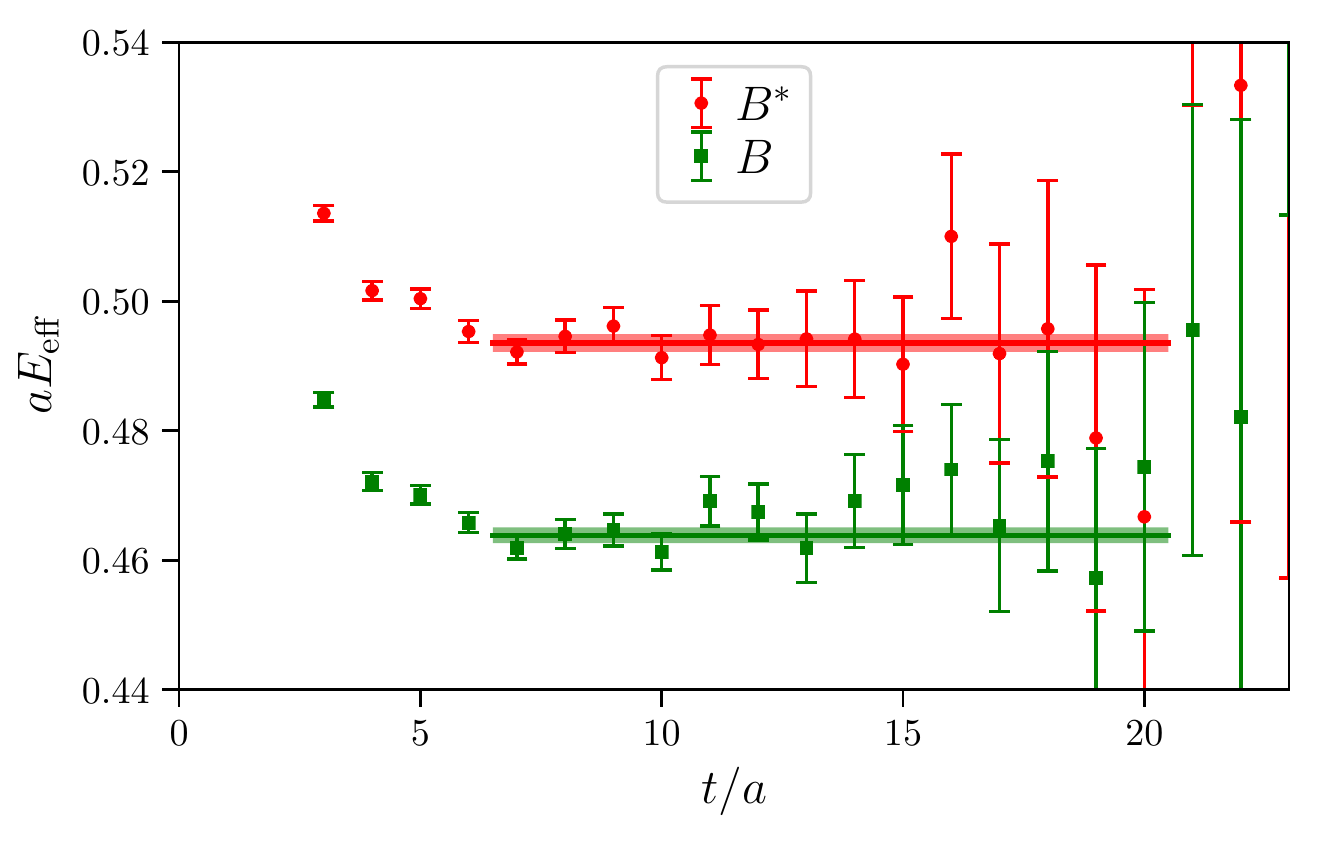}
	\includegraphics[width=0.45\linewidth]{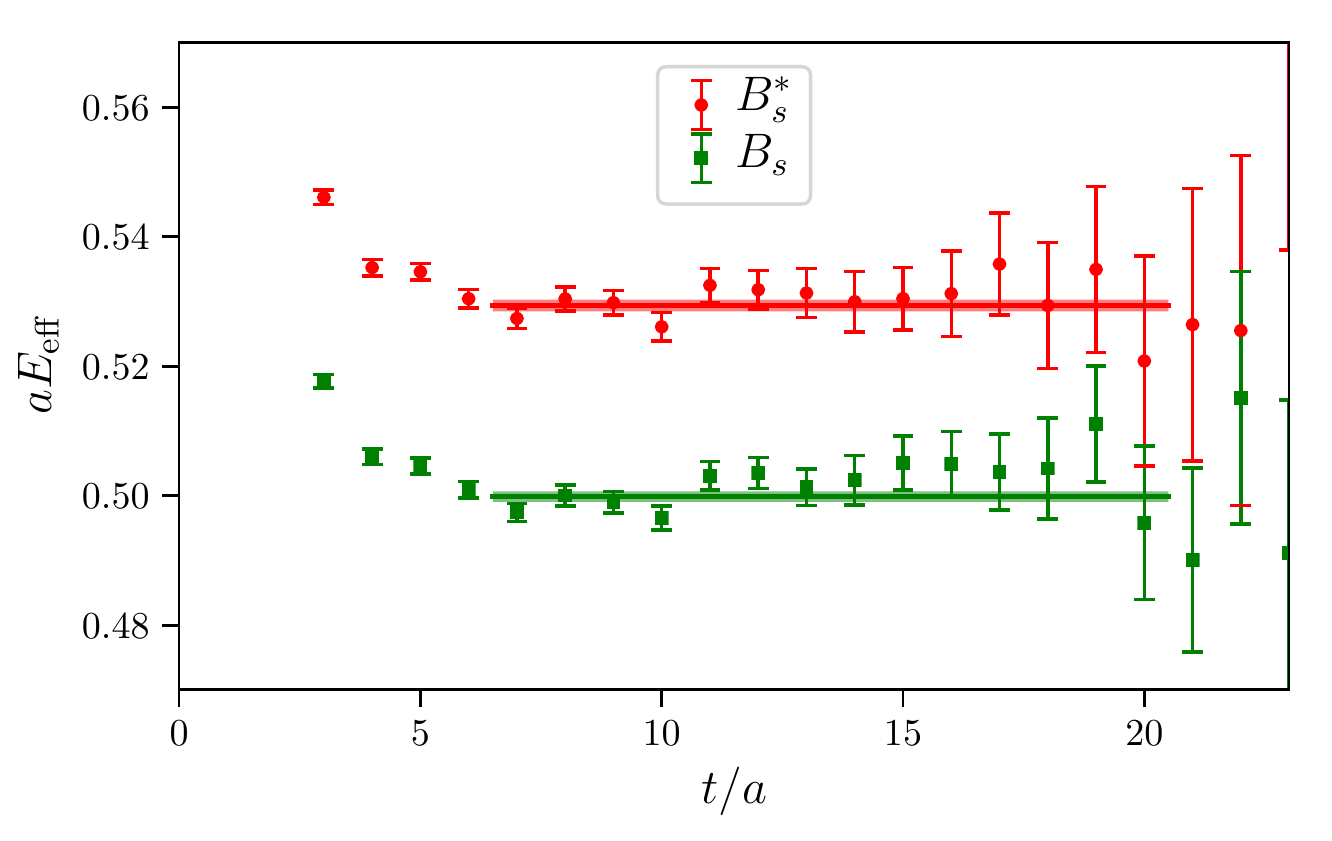}
	\includegraphics[width=0.45\linewidth]{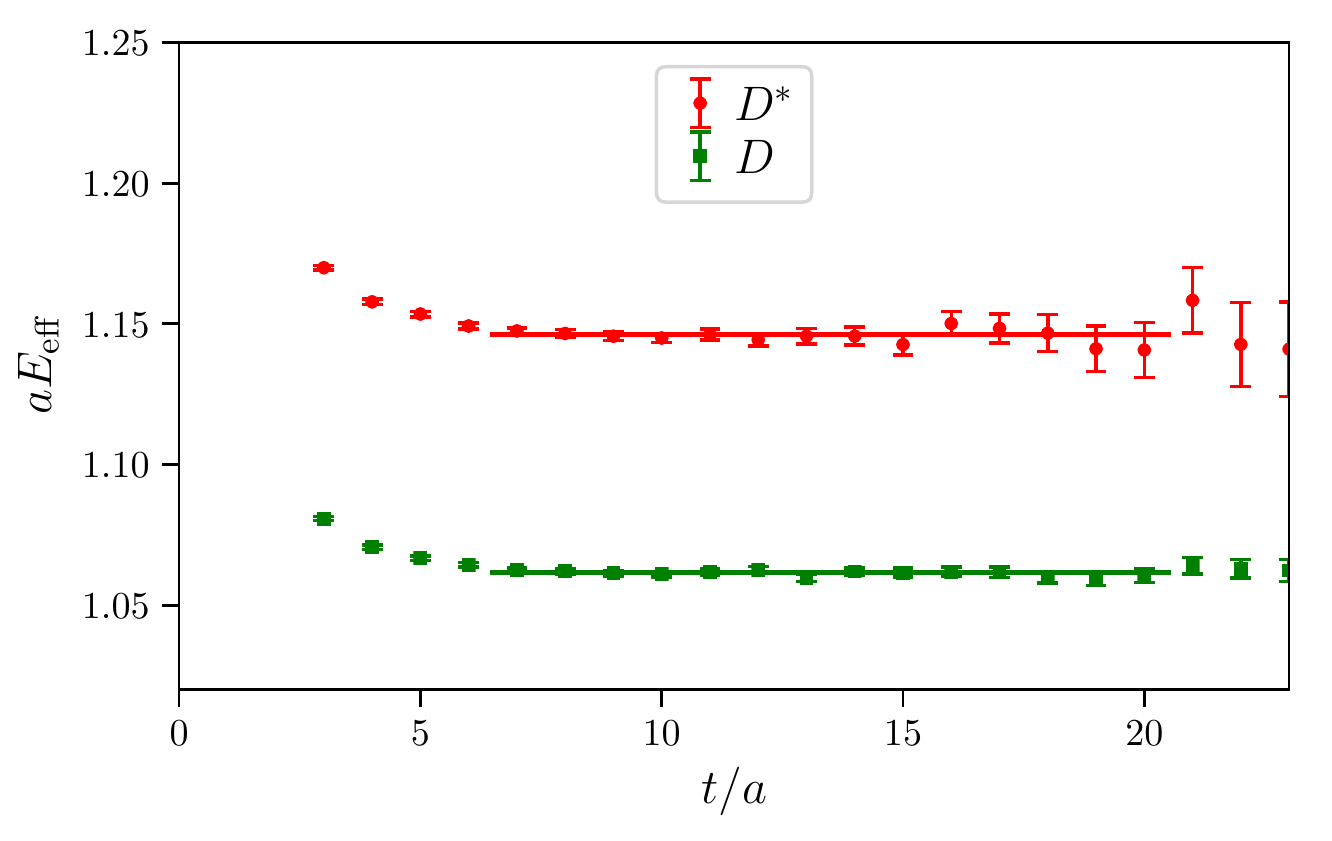}
	\caption{\label{fig:effm_plots_mesons}Effective energies for pseudoscalar and vector $ B $, $ B_s $ and $ D $ mesons for ensemble C005. The horizontal lines represent the corresponding plateau fits in the range $ t/a= 7 \dots 20 $.}
\end{figure}

We determined the ground-state energies of pseudoscalar and vector $ B $, $ B_s $ and $ D $ mesons via uncorrelated $\chi^2$-minimizing fits of constants to the corresponding effective-energy functions at sufficiently large temporal separations combined with a jackknife analysis. As usual, these effective energies are defined as
\begin{equation}
a E_{\textrm{eff}}(t) = \ln\bigg(\frac{C(t)}{C(t+a)}\bigg) ,
\end{equation}
where $ C(t) $ is a temporal correlation function of one of the interpolating operators (\ref{eq:Binterpolator}) to (\ref{eq:Dastinterpolator}). The results for all six mesons for each of the five ensembles are listed in \cref{tab:meson_energies}. As a cross-check we also determined these meson energies by correlated exponential fitting as in our previous work \cite{Leskovec:2019ioa} and found consistent results. To exemplify the quality of our numerical data, we show in \cref{fig:effm_plots_mesons} effective-energy plots for ensemble C005 together with the corresponding plateau fits.

\begin{table}[htb]
	\centering
	\begin{tabular}{ccccccc}\hline\hline
		Ensemble	& $ aE_{B} $	& $ aE_{B^*} $	& $ aE_{B_s} $	& $ aE_{B_s^*} $	& $ aE_{D} $	& $ aE_{D^*} $ \\ \hline
		C00078		& 0.4564(46) & 0.4814(49) & 0.5052(12) & 0.5349(15) & 1.0823(14) & 1.1638(21) \\
		C005		& 0.4639(12) & 0.4936(14) & 0.4998(8)  & 0.5294(9)  & 1.0616(4)  & 1.1462(8)  \\
		C01			& 0.4737(11) & 0.5052(13) & 0.5025(8)  & 0.5338(10) & 1.0714(4)  & 1.1586(7)  \\
		F004		& 0.3757(10) & 0.3976(11) & 0.4031(6)  & 0.4256(7)  & 0.7944(4)  & 0.8566(6)  \\
		F006		& 0.3786(6)  & 0.4007(7)  & 0.4033(4)  & 0.4258(5)  & 0.7981(2)  & 0.8609(4)  \\ \hline\hline
	\end{tabular}
\caption{\label{tab:meson_energies}Energies of pseudoscalar and vector $ B $, $ B_s $ and $ D $ mesons.}
\end{table}

Note that the energies of the $ B $, $ B^\ast $, $ B_s $ and $ B^\ast_s $ mesons listed in Table~\ref{tab:meson_energies} do not correspond to the full meson masses, as e.g.\ measured in experiment. The reason is the use of NRQCD, resulting in negative energy shifts proportional to $n_b$, the number of $b$ quarks present in the corresponding states. At tree level, this shift amounts to $-n_b m_b$, where $m_b$ is the $b$-quark mass. Since we exclusively consider energy differences between four-quark states and meson-meson thresholds with the same $n_b$, these energy shifts cancel and there is no need to determine them.

% ********************
% ********************
% ********************
% ********************
% ********************

\section{Results on antiheavy-antiheavy-light-light four-quark systems \label{sec:results}}

The correlation matrix (\ref{eq:defCorrelationMatrix}) with interpolating operators from \cref{sec:fourQuarkSystems_bbus}, \cref{sec:ops_bcudJ0} or \cref{sec:ops_bcudJ1} can be written as a sum over the energy eigenstates $ |n\rangle $ of the respective flavor and $J^P$ sector,
\begin{equation}
C_{jk}(t) = \sum_{n=0}^{\infty} Z_j^n Z_k^n \textrm{e}^{-E_nt} ,
\label{eq:CorrelationMatrix}
\end{equation}
with real valued
\begin{equation}
Z_j^n =\langle \Omega | \op_j | n\rangle
\end{equation}
and $ |\Omega\rangle $ denoting the vacuum. To extract the energy levels $E_n$ and overlap factors $Z_j^n$ from the numerical lattice-QCD results for $C_{jk}(t)$, we carry out correlated $\chi^2$-minimizing multi-exponential fits of a truncated version of the right hand side of \cref{eq:CorrelationMatrix},
\begin{equation}
C^\text{fit}_{jk}(t) = \sum_{n=0}^{N-1} Z_j^n Z_k^n \textrm{e}^{-E_nt} ,
\label{eq:CorrelationMatrix_truncated}
\end{equation}
in a suitably chosen range $ t_{\textrm{min}} \le t \le t_{\textrm{max}} $. For further technical details concerning this multi-exponential fitting we refer to Section~V~A of our previous work \cite{Leskovec:2019ioa}. To check for and to exclude systematic errors as well as to minimize statistical errors, we also consider submatrices of the correlation matrices defined in \cref{SEC597} and vary the temporal fit range.

% ********************

\subsection{$ \bar{b}\bar{b}us $ with $ J^P=1^+ $}

% **********

\subsubsection{Reduction of the size of the correlation matrix from $7 \times 4$ to $6 \times 3$}

\begin{figure}[htb]
	\centering
	\includegraphics[width=0.49\linewidth, page=1, trim = 30 12 50 25 ,clip]{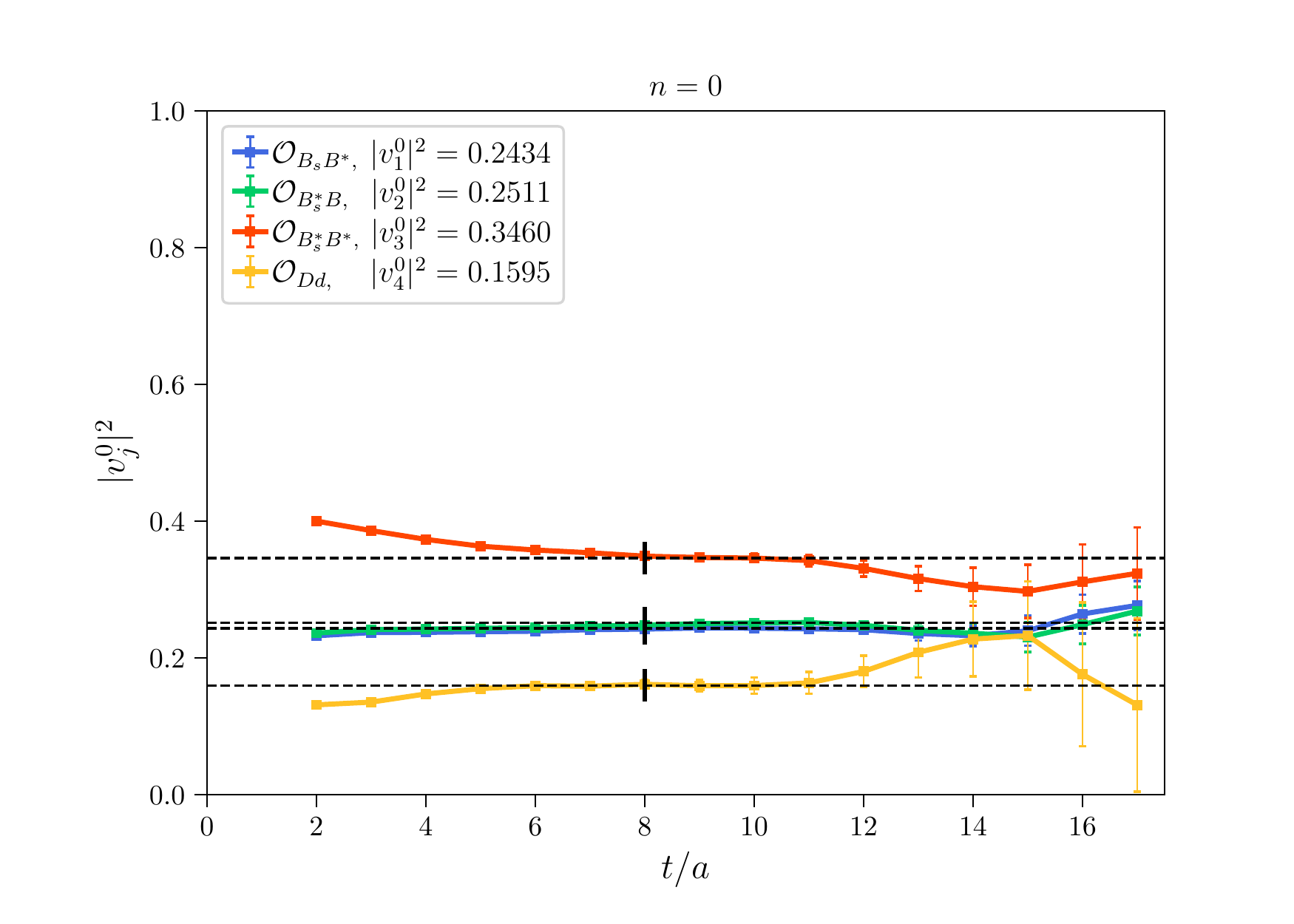}
	\includegraphics[width=0.49\linewidth, page=2, trim = 30 12 50 25 ,clip]{./figs/EV_bbus_C01_components.pdf}
	
	\includegraphics[width=0.49\linewidth, page=3, trim = 30 12 50 25 ,clip]{./figs/EV_bbus_C01_components.pdf}
	\includegraphics[width=0.49\linewidth, page=4, trim = 30 12 50 25 ,clip]{./figs/EV_bbus_C01_components.pdf}
	
	\caption{\label{fig:EvComponents}Squared normalized eigenvector components $|v_j^n|^2$ as functions of $t$ for ensemble~C01, obtained by solving generalized eigenvalue problems as defined in Eq.\ (\ref{EQN533}). The corresponding $4 \times 4$ correlation matrix contains the four local interpolating operators (\ref{eq:op_BBsast_total_zero}) to (\ref{eq:op_Dd_total_zero}). The dashed horizontal lines represent the squares of the coefficients $\bar{v}_j^n$, where $\bar{v}_j^n = v_k^n(t/a = 8)$.}
\end{figure}

In a preparatory step we replace the local interpolating operators (\ref{eq:op_BBsast_total_zero}) to (\ref{eq:op_Dd_total_zero}) by linear combinations of these operators,
\begin{equation}
\op'_j = \sum_{k=1}^4 \bar{v}_k^{j-1} \op_k \quad , \quad j = 1,\ldots,4 .
\label{eq:lin_comb_EV_operator}
\end{equation}
The coefficients $\bar{v}_j^n$ were determined by solving generalized eigenvalue problems
\begin{equation}
\label{EQN533} \sum_{k=1}^4 C_{jk}(t) v_k^n(t) = \lambda^n(t) \sum_{k=1}^4 C_{jk}(t_0=a) v_k^n(t) \quad , \quad j = 1,\ldots,4 \quad , \quad n = 0,\ldots,3 ,
\end{equation}
where $C_{jk}(t)$ is the lattice-QCD result for the $4 \times 4$ correlation matrix containing the local operators $\op_1$, $\op_2$, $\op_3$ and $\op_4$. We normalized the eigenvector components such that $\sum_j |v_j^n(t)|^2 = 1$ and show them for ensemble C01 in \cref{fig:EvComponents}, where one can see that the eigenvector components $v_j^n(t)$ are fairly independent of $t$, in particular for larger values of $t$. Thus we defined the coefficients in \cref{eq:lin_comb_EV_operator} as $\bar{v}_j^n = v_k^n(t/a = 8)$, where $t/a = 8$ was selected because the $v_k^n(t/a = 8)$ have rather small statistical uncertainties and are already consistent with the plateaus formed at larger values of $t$ (for ensemble C01 the coefficients $\bar{v}_j^n$ are collected in \cref{tab:ev_components_C01}; for the other four ensembles they are quite similar). With this definition, operator $\op'_j$, when applied to the vacuum, should create a trial state with large overlap to energy eigenstate $ | j-1 \rangle $. Thus, this new set of operators offers the possibility to discard some of them (e.g.\ $\op'_4$ or even $\op'_4$ and $\op'_3$) to keep the corresponding correlation matrix small, while retaining at the same time the overlap to the low-lying energy eigenstates of interest. This is beneficial for the precision of the numerical analyses discussed below.

\begin{table}[htb]
	\centering
	\begin{tabular}{C{1.5cm}C{1.5cm}C{1.5cm}C{1.5cm}C{1.5cm}}\hline\hline
		$ \bar{v}_j^n $	&  $j=1 $	& $ j=2 $ 	& $ j=3 $ & $ j=4 $ \\ \hline 
		$ n=0 $		& $ +0.493 $ & $ -0.501 $	& $ -0.588 $ & $ -0.399 $ \\ 	
		$ n=1 $		& $ -0.708 $ & $ -0.706 $ & $ +0.002 $ & $ +0.002 $ \\ 
		$ n=2 $		& $ -0.448 $ & $ +0.446 $ & $ -0.773 $ & $ -0.056 $ \\ 
		$ n=3 $		& $ -0.351 $ & $ +0.351 $ & $ +0.529 $ & $ -0.689 $ \\ \hline\hline
	\end{tabular}
	\caption{\label{tab:ev_components_C01}Coefficients $\bar{v}_j^n$ defining the interpolating operators $\op'_j$ for ensemble C01 [see Eq.\ (\ref{eq:lin_comb_EV_operator})].}
\end{table}

Since we are mainly interested in the energy level of the ground state, $\op_1'$ is of particular importance. In practice, it turned out that using in addition also $\op_2'$ and $\op_3'$ is favorable with respect to a precise determination of energy levels. $\op_4'$, however, does not seem to be advantageous in our context and is therefore discarded. Altogether our analysis is based on the three local interpolating operators $\op_1'$, $\op_2'$, $\op_3'$ and the three non-local interpolating operators
\begin{equation}
\op'_4 = \op_5 , \qquad \op'_5 = \op_6 , \qquad \op'_6 = \op_7
\end{equation}
defined in Eqs.\ (\ref{eq:op_BBsast_sep_zero}) to (\ref{eq:op_BastBsast_sep_zero}). Thus, in the following we will study a $ 6 \times 3 $ correlation matrix and its submatrices.

% **********

\subsubsection{\label{SEC562}Energy levels}

\begin{figure}[!htb]
	\centering
	\includegraphics[width=0.95\linewidth,  trim = 0 0 0 0 ,clip]{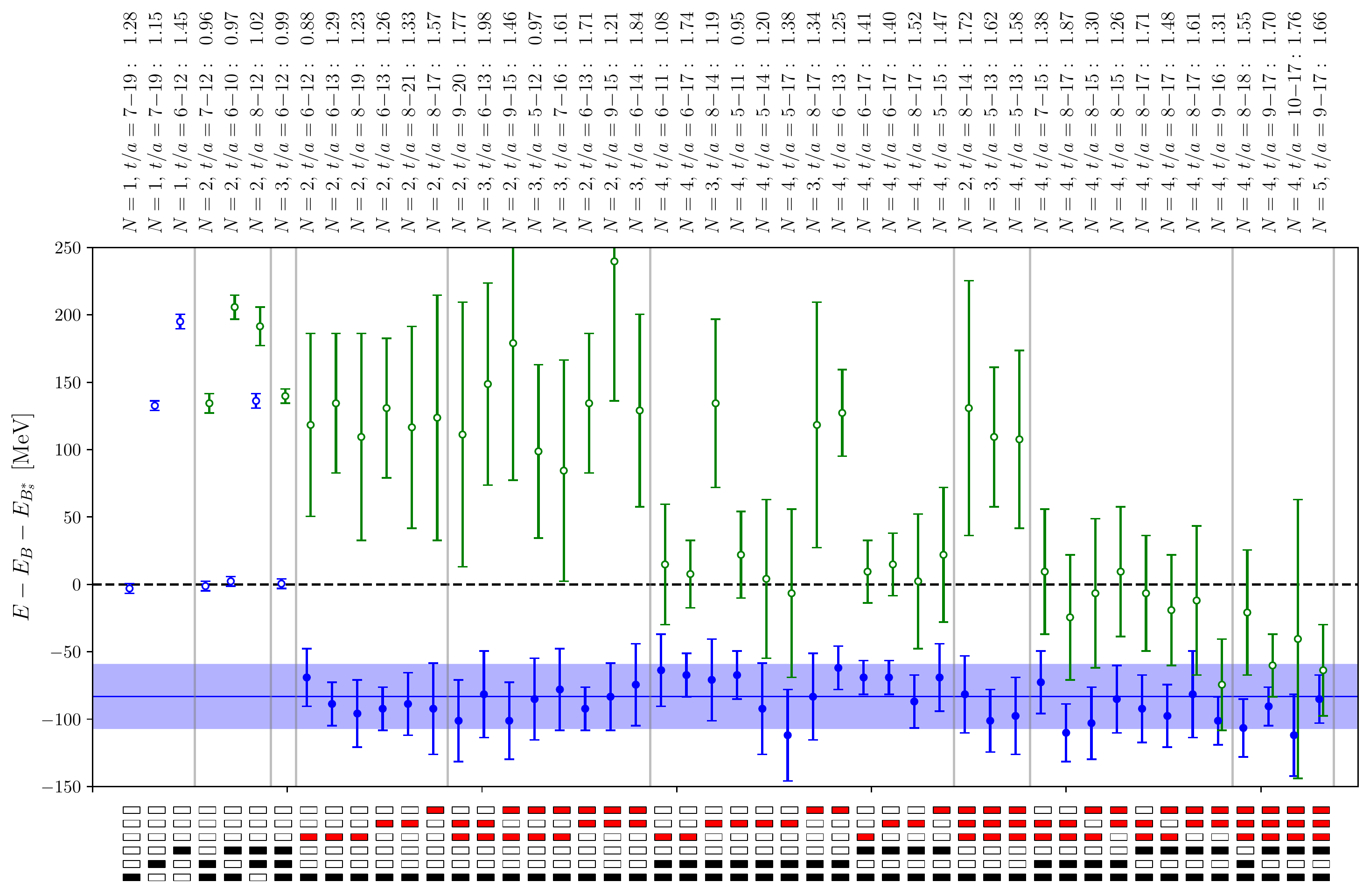}
	\caption{\label{fig:fit_results_bbus_C01} Fit results for $E_0$ (blue) and $E_1$ (green) for the $ \bar{b}\bar{b}us $ system relative to the $ B B_s^\ast $ threshold for ensemble C01.}
\end{figure}

To reliably determine the lowest energy levels, in particular that of the ground state, we carried out multi-exponential fits as discussed at the beginning of this section. We considered various submatrices, numbers of exponentials $N$, and fit ranges $ t_{\textrm{min}} \le t \le t_{\textrm{max}} $. The corresponding results with correlated $\chi^2/\text{d.o.f.} < 2$ are summarized for ensemble C01 in \cref{fig:fit_results_bbus_C01}, while those for the other ensembles are collected in \cref{app:energytablesandfigures}. The boxes at the bottom of \cref{fig:fit_results_bbus_C01} indicate, for each fit, which interpolating operators were included. A filled/empty box represents an operator that was included/excluded. From bottom to top, the boxes represent $ \op'_1 $, $ \op'_2 $, ..., $ \op'_6 $. Local operators are colored in black, scattering operators in red. The fit results for $E_0$ and $E_1$ are shown as blue and green points with error bars, where the energy of the lowest threshold, $E_B + E_{B_s^\ast}$, is subtracted (this threshold is represented by the horizontal dashed line). Above the plot, further details are provided for each fit: the number of exponentials, the temporal fit range, and the resulting correlated $ \chi^2/\text{d.o.f.}\,. $

The first seven columns from the left represent fits in which only local interpolating operators were considered. Each of the three local operators seems to be associated with a specific energy, $ \op'_1 $ with $\approx 0 \, \text{MeV}$, $ \op'_2 $ with $\approx 130 \, \text{MeV}$ and $ \op'_3 $ with $\approx 200 \, \text{MeV}$. This is not surprising, given that these operators were constructed in such a way that the corresponding $3 \times 3$ correlation matrix is approximately diagonal in the region of $t$ separations that enter the multi-exponential fits. Clearly, $ \op'_1 $ is of particular importance for a precise determination of the energy of the ground state. Thus, $ \op'_1 $ was included in all further fits, where in addition to local operators also scattering operators were used.

It is crucial to note that for all fits that include at least $ \op'_1 $ and one of the scattering operators $ \op'_4 $ to $ \op'_6 $, the fit result for $E_0$ is around $100 \, \text{MeV}$ below the $ BB_s^\ast $ threshold. This is a first clear indication that the ground state in the $ \bar{b}\bar{b}us $ and $ J^P=1^+ $ sector is a strong-interaction-stable tetraquark. One can also see that the fit result for $E_1$ is in many cases close to $0 \, \text{MeV}$, which is consistent with the expectation that the first excitation is a meson-meson scattering state close to the $ BB_s^\ast $ threshold. We note that the results for the other four ensembles are comparable, i.e.\ $E_0$ is around $100 \, \text{MeV}$ below the $ BB_s^\ast $ threshold, and $E_1$ is  around $0 \, \text{MeV}$ for several fits (see \cref{app:energytablesandfigures}).

As one can see from \cref{fig:fit_results_bbus_C01}, the results for $E_0$ from fits including at least $ \op'_1 $ and one of the scattering operators $ \op'_4 $ to $ \op'_6 $ (represented by the filled blue data points) agree within the statistical uncertainties. Thus, these fit results seem to be suited to estimate the ground state energy and its uncertainty. We computed such an estimate by a weighted average of these fit results, assuming 100\% correlation, using a standard method also employed by the FLAG collaboration \cite{FlavourLatticeAveragingGroup:2019iem} (see \cref{SEC607} for a brief summary). The estimated ground-state energies are also plotted in the corresponding figures, e.g.\ for ensemble C01 in \cref{fig:fit_results_bbus_C01} (the blue horizontal line and the light blue error band).

Concerning the energy of the first excitation, \cref{fig:fit_results_bbus_C01} suggests that it is somewhere around the $ BB_s^\ast $ threshold. We refrain from estimating this energy in a quantitative way by computing a weighted average of selected fit results for $E_1$. The reason is that it is hard to decide whether $E_1$ obtained by a particular fit indeed corresponds to the energy of the first excitation. There are several states that could be close to the $ B B_s^\ast $ threshold, e.g.\ a $ B B_s^\ast $ or a $ B^\ast B_s $ scattering state. Additionally, there might also be a $ B^\ast B_s^\ast $ scattering state in that energy region because of the finite spatial volume and the attractive interaction of the two mesons \cite{Bicudo:2016ooe}. The low-lying excitations could correspond to superpositions of these structures and are expected to have similar energies. Thus, a fit result for $E_1$ close to the $ BB_s^\ast $ threshold could, for example, reflect the energy of the first or the second excitation or a mix of both. In principle, one could try to disentangle these excitations by studying the resulting overlap factors $Z_j^n$ for each fit in detail. Since we only need the ground-state energy for our final analysis in \cref{SEC489}, we discuss the overlap factors just for a single fit with $N = 3$ exponentials to the full $6 \times 3$ correlation matrix (see the following subsection).

% **********

\subsubsection{Overlap factors}

A trial state $ \op'_j{}\dagg | \Omega \rangle $ can be expanded according to
\begin{equation}
\op'_j{}\dagg | \Omega \rangle  = \sum_{n=0}^\infty | n \rangle \langle n | \op'_j{}\dagg | \Omega \rangle = \sum_{n=0}^\infty Z_j^n | n \rangle ,
\end{equation}
which shows that the overlap factors $Z_j^n$ contain information about the composition and quark arrangement of the energy eigenstates $ |n\rangle $. For example, an overlap factor $ |Z_j^n| $ that is significantly larger than all other overlap factors $ |Z_j^m| $ with $m \neq n$ indicates that the trial state $\op'_j{}\dagg | \Omega \rangle$ is quite similar to the eigenstate $| n \rangle$. Vice versa, if the overlap factor $ |Z_j^n| $ is significantly smaller than at least one of the other overlap factors $ |Z_j^m| $ with $m \neq n$, one can conclude that the trial state $\op'_j{}\dagg | \Omega \rangle$ is almost orthogonal to the eigenstate $| n \rangle$.

\begin{figure}[!htb]
	\centering
	\includegraphics[width=0.95\linewidth]{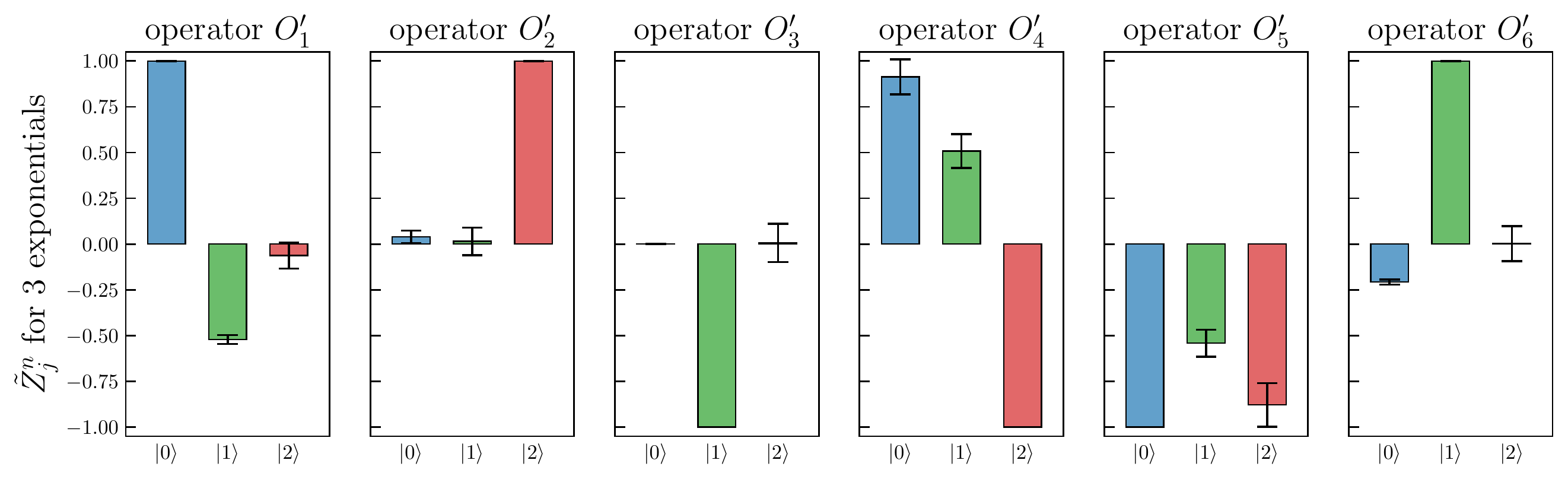}
	\caption{\label{fig:overlap_factors_bbus}Normalized overlap factors $ \tilde{Z}_j^n $ for the $ \bar{b}\bar{b}us $ system obtained via a multi-exponential fit with $N = 3$ in the range $16 \leq t/a \leq 24$ to the full $6 \times 3$ correlation matrix of ensemble F004. The index of the operator above each plot is identical to the index $j$, while the labels of the energy eigenstates below each plot correspond to the index $n$.}
\end{figure}

In \cref{fig:overlap_factors_bbus} we show normalized overlap factors
\begin{equation}
\tilde{Z}_j^n = \frac{Z_j^n}{\textrm{max}_m(|Z_j^m|)}
\end{equation}
obtained via a multi-exponential fit with $N = 3$ in the range $16 \leq t/a \leq 24$ to the full $6 \times 3$ correlation matrix of ensemble F004. Corresponding results for the other ensembles are qualitatively identical. We start with an extensive discussion of the overlap factors $Z_j^0$ associated with the ground state $| 0 \rangle$ and then briefly comment on the overlap factors $Z_j^n$ with $n > 0$ related to the excitations.

The result $|Z_1^0| \gg |Z_1^1| , |Z_1^2|$ suggests that the trial state $\op'_1{}\dagg | \Omega \rangle$ has a large ground-state overlap, i.e., is rather similar to the ground state. Recall that $\op'_1$ is a weighted sum of four local operators $\op_1$ to $\op_4$ (\cref{eq:lin_comb_EV_operator} with coefficients $\bar{v}_j^0$ as listed in \cref{tab:ev_components_C01} for ensemble C01). Since $\bar{v}_1^0 \approx -\bar{v}_2^0$, there is a local $B B_s^\ast$ and $B^\ast B_s$ component (operators $\op_1$ and $\op_2$) that is antisymmetric in the light flavors $u s$. There is also a local antisymmetric $B^\ast B_s^\ast$ component (operator $\op_3$) of the same order of magnitude. Such a meson-meson composition is expected from existing static-light lattice QCD results \cite{Bicudo:2016ooe} on the strong-interaction-stable $\bar b \bar b u d $ tetraquark with $I(J^P) = 0(1^+)$, a closely related four-quark system (same quantum numbers $J^P$, and because of isospin $I = 0$ antisymmetric in the light flavors), where it was found that it is a roughly even mixture of $B B^\ast$ and $B^\ast B^\ast$. The $\bar b \bar b u s$ system  also has a sizable diquark-antidiquark component (operator $\op_4$), albeit somewhat smaller than the aforementioned meson-meson components. This, too, is expected and is consistent with recent static-light lattice-QCD results on the $\bar b \bar b u d $ tetraquark, where the meson-meson to diquark-antidiquark ratio was estimated to be around $60\% / 40\%$ \cite{Bicudo:2021qxj}.

The overlap factors $Z_2^n$ and $Z_3^n$ clearly show that the trial states $\op'_2{}\dagg | \Omega \rangle$ and $\op'_3{}\dagg | \Omega \rangle$ are essentially orthogonal to the ground state $| 0 \rangle$. According to \cref{tab:ev_components_C01}, the operator $\op'_2$ is a local combination of $B B_s^\ast$ and $B^\ast B_s$ (operators $\op_1$ and $\op_2$) that is \emph{symmetric} in the light flavors $u s$, i.e.\ the analog of an $I = 1$ operator for light flavors $u d$. This confirms that the $ \bar b \bar b u s $ ground state is antisymmetric in the light flavors and indicates that it is the counterpart of the $\bar b \bar b u d $ tetraquark with $I(J^P) = 0(1^+)$. While the operator $\op'_3$ is flavor antisymmetric, it was constructed via the GEVP in a way to generate almost no overlap with the ground state and with the lowest flavor-symmetric excitation $|2\rangle$. Thus it is not surprising that $\tilde{Z}_3^0 \approx \tilde{Z}_3^2 \approx 0$.

The scattering trial states $\op'_4{}\dagg | \Omega \rangle$ and $\op'_5{}\dagg | \Omega \rangle$ both have overlaps to the ground state $| 0 \rangle$, but also sizable overlaps to the first and second excitations. Thus, one should not infer that the ground state is quite similar to a scattering state. Since the scattering operators $\op'_4$ and $\op'_5$ contain all terms present in the local operators $\op_1$ and $\op_2$, the non-vanishing overlaps $Z_4^0$ and $Z_5^0$ rather support our conclusions above, namely that the $\bar b \bar b u s $ ground state is a four-quark bound state with a large local flavor antisymmetric $B B_s^\ast$ and $B^\ast B_s$ component.

As already discussed above in the context of energy levels and the fit parameter $E_1$, one should be cautious in formulating conclusions concerning the excitations based on our multi-exponential fit results. Still, it seems noteworthy to mention that the trial states $\op'_3{}\dagg | \Omega \rangle$ and $\op'_6{}\dagg | \Omega \rangle$ have large overlap with the first excitation $| 1 \rangle$ and only little overlap with $| 0 \rangle$ and $| 2 \rangle$. Since $\op'_6$ is a $B^\ast B_s^\ast$ scattering operator and the dominant component of $\op'_3$ is a local $B^\ast B_s^\ast$ structure (see \cref{tab:ev_components_C01}), this might be a hint that the first excitation is of $B^\ast B_s^\ast$ type or at least contains a significant $B^\ast B_s^\ast$ component. Even though the $B^\ast B_s^\ast$ threshold is around $50 \, \text{MeV}$ above the $B B_s^\ast$ threshold, the expected attraction of a $B^\ast$ meson and a $B_s^\ast$ meson (see Ref.\ \cite{Bicudo:2016ooe}) and the finite spatial volume could lead to an energy level of the first excitation close to the $B B_s^\ast$ threshold, as indicated by \cref{fig:fit_results_bbus_C01}.

Finally, the overlap factors $Z_j^2$ represent almost exclusively symmetric light flavor combinations. This indicates that also for the scattering states in our finite spatial lattice volume, SU(3) flavor symmetry is approximately preserved. Thus, the second excitation seems to be the analog of the ground state in the $\bar b \bar b u d$ four-quark sector with $I = 1$, where no strong-interaction-stable four-quark state was found in a static-light lattice-QCD study \cite{Bicudo:2015kna,Bicudo:2016ooe}.

% ********************

\subsection{$ \bar{b}\bar{c}ud $ with $ I(J^P)=0(0^+) $}

As discussed in \cref{sec:ops_bcudJ0}, we consider three interpolating operators for this system: two local operators and one scattering operator. Thus, the corresponding correlation matrix has size $3 \times 2$. Since this is a rather small matrix, there is no need to further reduce the number of operators in a preparatory step, as done for the $ \bar b \bar b u s $ system.

\begin{figure}[!htb]
	\centering
	\includegraphics[width=0.55\linewidth, trim = 0 0 0 0 ,clip]{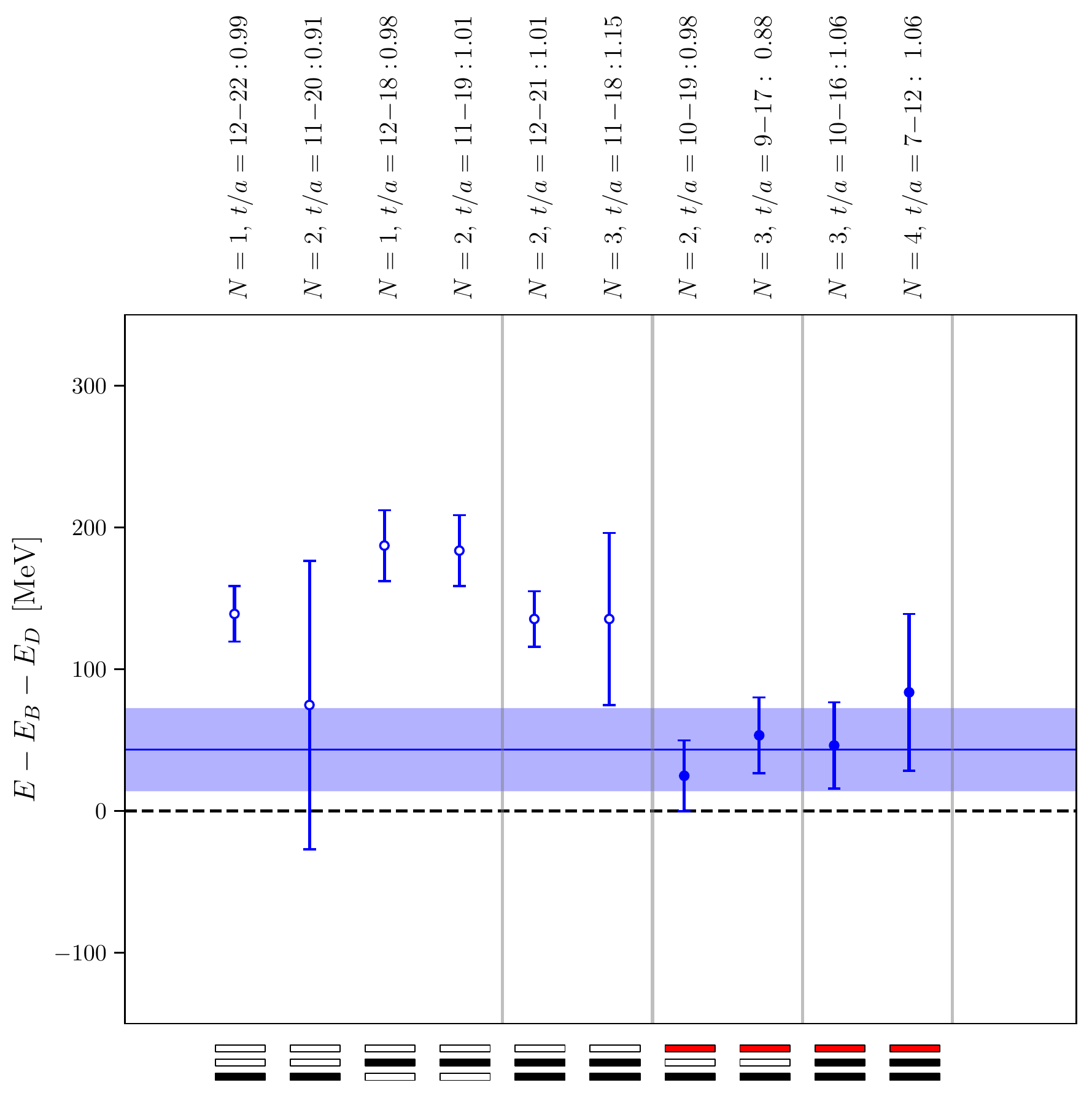}
	\caption{\label{fig:fit_results_bcudJ0_C01} Fit results for $E_0$ for the $ \bar{b}\bar{c}ud $ system with $ I(J^P)=0(0^+) $ relative to the $ B D $ threshold for ensemble C01.}
\end{figure}

To determine the energy of the ground state we proceed as in \cref{sec:ops_bcudJ0} and carry out multi-exponential fits. Again we consider various submatrices, numbers of exponentials $N$, and fit ranges $ t_{\textrm{min}} \le t \le t_{\textrm{max}} $. The corresponding results with correlated $\chi^2/\text{d.o.f.} < 2$ are summarized for ensemble C01 in \cref{fig:fit_results_bcudJ0_C01}, while those for the other ensembles are collected in \cref{app:energytablesandfigures}.

Like for the $ \bar b \bar b u s $ system, we again find significantly lower values for $E_0$ once the scattering operator $\op_3$ (see \cref{eq:op_BD_sep_zero}) is included, compared to fits in which only local interpolating operators are used.
Averaging over the fits that include the scattering operator leads to an estimate for the ground-state energy, which is slightly above, but within its uncertainty compatible with, the $ B D $ threshold.
We find similar results for the other four ensembles (see \cref{app:energytablesandfigures}).
This suggest that there is no strong-interaction-stable four-quark state in this channel. The lowest energy eigenstate rather seems to be a $ B D $ scattering state.

\begin{figure}[!htb]
	\centering
	\includegraphics[width=0.5\linewidth]{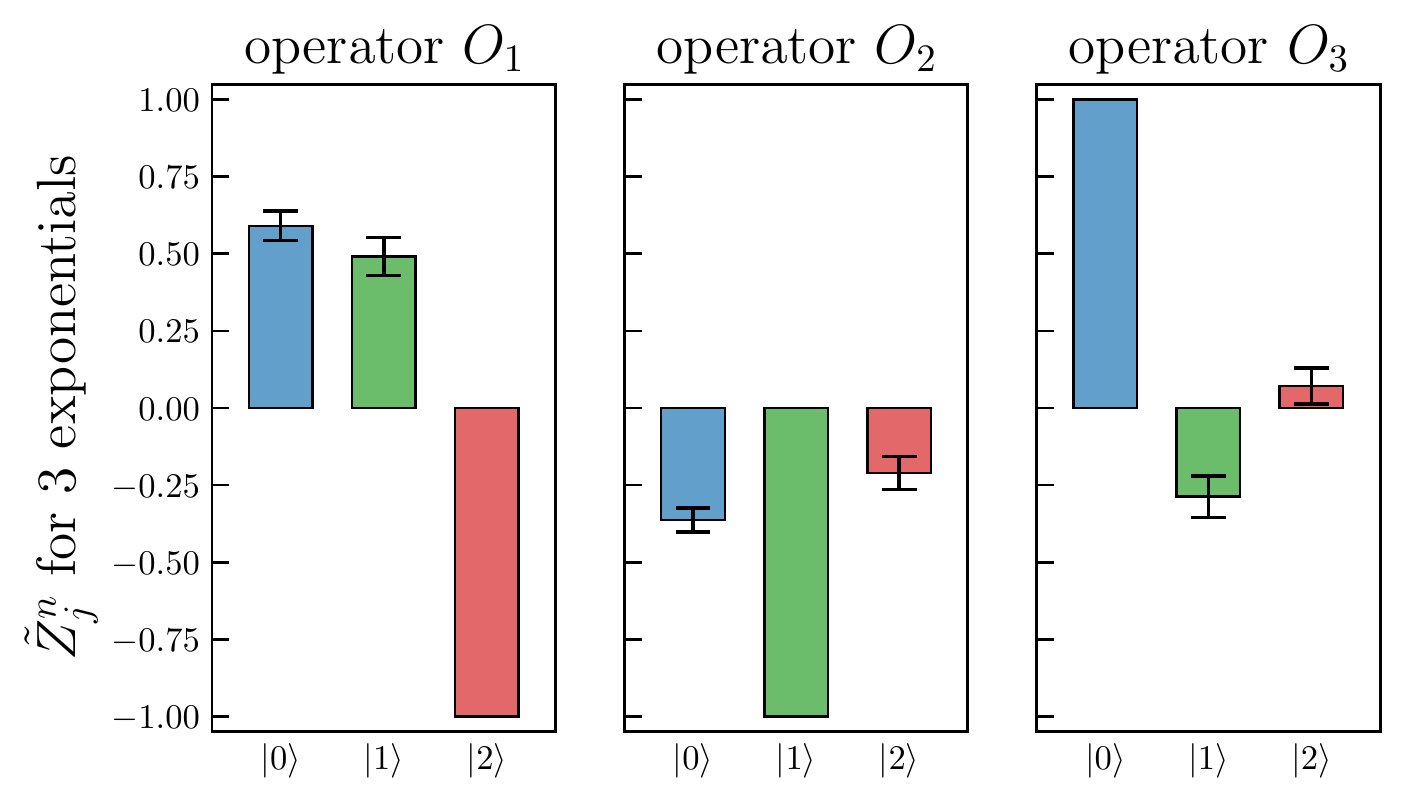}
	\caption{\label{fig:overlap_factors_bcudJ0}Normalized overlap factors $ \tilde{Z}_j^n $ for the $ \bar{b}\bar{c}ud $ system with $ I(J^P)=0(0^+) $ obtained via a multi-exponential fit with $N = 3$ in the range $6 \leq t/a \leq 10$ to the full $3 \times 2$ correlation matrix of ensemble F004. The index of the operator above each plot is identical to the index $j$, while the labels of the energy eigenstates below each plot correspond to the index $n$.}
\end{figure}

In \cref{fig:overlap_factors_bcudJ0} we show the normalized overlap factors $ \tilde{Z}_j^n $ obtained via a multi-exponential fit with $N = 3$ in the range $6 \leq t/a \leq 10$ to the full $3 \times 2$ correlation matrix of ensemble F004. Corresponding results for the other ensembles are qualitatively identical.
It is obvious that the $ B D $ scattering trial state $\op_3{}\dagg | \Omega \rangle$ has large overlap to the ground state and almost negligible overlap to the first and second excitation. This supports our above conclusion that the ground state is a meson-meson scattering state.

% ********************

\subsection{\label{sec:resultsbcudJ1}$ \bar{b}\bar{c}ud $ with $ I(J^P)=0(1^+) $}

According to \cref{sec:ops_bcudJ1} we consider five interpolating operators here: three local operators and two scattering operators. Thus, the corresponding correlation matrix has size $5 \times 3$. We do not reduce the number of operators in a preparatory step as done for the $ \bar b \bar b u s $ system.

To determine the energies of the ground state and of the first excitation, we again carry out multi-exponential fits and consider various submatrices, numbers of exponentials $N$, and fit ranges $ t_{\textrm{min}} \le t \le t_{\textrm{max}} $. The corresponding results with correlated $\chi^2/\text{d.o.f.} < 2$ are summarized for ensemble C01 in \cref{fig:fit_results_bbus_C01}, while those for the other ensembles are collected in \cref{app:energytablesandfigures}.

\begin{figure}[!htb]
	\centering
	\includegraphics[width=0.7\linewidth, trim = 0 0 0 0 ,clip]{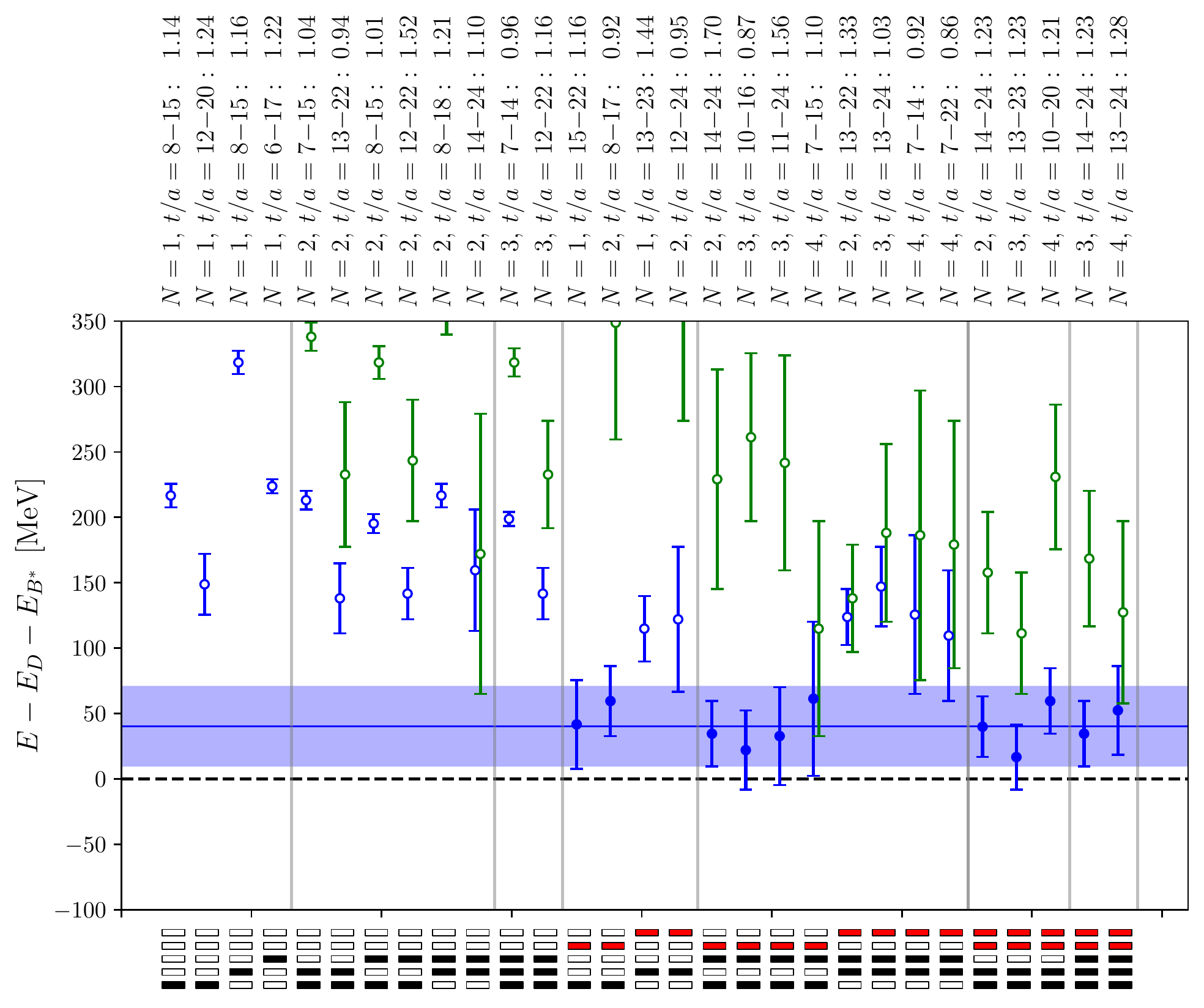}
	\caption{\label{fig:fit_results_bcudJ1_C01} Fit results for $E_0$ (blue) and $E_1$ (green) for the $ \bar{b}\bar{c}ud $ system with $ I(J^P)=0(1^+) $ relative to the $ B^\ast D $ threshold for ensemble C01.}
\end{figure}

As for the previously investigated four-quark systems, we find significantly lower values for $E_0$ and $E_1$ as soon as the scattering operators $\op_4$ and $\op_5$ (see \cref{eq:op_BastD_sep_zero} and \cref{eq:op_BDast_sep_zero}) are included. In particular, operator $ \op_4 $, which has a $ B^\ast D $-like meson-meson structure, favors small values for $E_0$ close to the $ B^\ast D $ threshold. Since the local operator $ \op_1 $ is also of $ B^\ast D $ type, we estimate the ground state energy by averaging over the fits that include both $ \op_1 $ and $ \op_4 $. The result is slightly above, but within its uncertainty compatible with, the $ B^\ast D $ threshold. As before, we do not estimate the energy of the first excitation quantitatively by computing a weighted average of selected fit results for $E_1$. We note, however, that this energy level seems to be close to the $ B D^\ast $ threshold, which is around $100 \, \text{MeV}$ above the $ B^\ast D $ threshold. We found similar results for the other four ensembles (see \cref{app:energytablesandfigures}). In summary, this suggests that there is no strong-interaction-stable four-quark state in this channel. The lowest energy eigenstate rather seems to be a $ B^\ast D $ scattering state.

\begin{figure}[!htb]
	\centering
	\includegraphics[width=0.9\linewidth]{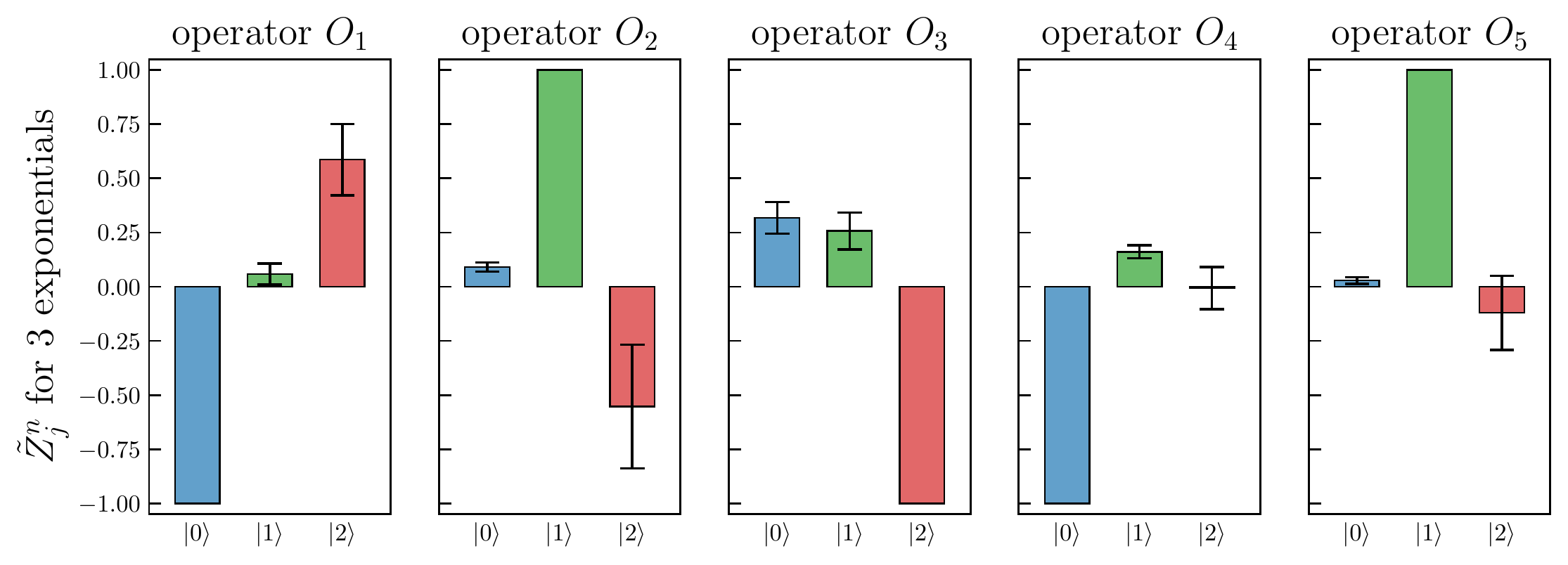}
	\caption{\label{fig:overlap_factors_bcudJ1}Normalized overlap factors $ \tilde{Z}_j^n $ for the $ \bar{b}\bar{c}ud $ system with $ I(J^P)=0(1^+) $ obtained via a multi-exponential fit with $N = 3$ in the range $14 \leq t/a \leq 20$ to the full $5 \times 3$ correlation matrix of ensemble F004. The index of the operator above each plot is identical to the index $j$, while the labels of the energy eigenstates below each plot correspond to the index $n$.}
\end{figure}

In \cref{fig:overlap_factors_bcudJ1} we show the normalized overlap factors $ \tilde{Z}_j^n $ obtained via a multi-exponential fit with $N = 3$ in the range $14 \leq t/a \leq 20$ to the full $5 \times 3$ correlation matrix of ensemble F004. Corresponding results for the other ensembles are qualitatively identical.
One can see that the $ B^\ast D $ scattering trial state $\op_4{}\dagg | \Omega \rangle$ almost exclusively overlaps with the ground state, i.e.\ $Z_4^0 \gg Z_4^1, Z_4^2$. Similarly, $Z_5^1 \gg Z_5^0, Z_5^2$, i.e.\ the $ B D^\ast $ scattering trial state $\op_5{}\dagg | \Omega \rangle$  almost exclusively overlaps with the first excitation. This supports our interpretation of the ground state and the first excitation as $ B^\ast D $ and $ B D^\ast$ scattering states.

% ********************

\subsection{\label{SEC489}Final results for the $ \bar{b}\bar{b}us $ and  $ \bar{b}\bar{c}ud $ ground-state energies}

We list the final results for the ground-state energies relative to the lowest meson-meson thresholds for the three investigated four-quark systems and for all five ensembles in \cref{tab:finalResults}. These energies correspond to the horizontal blue lines and light blue error bands in \cref{fig:fit_results_bbus_C01}, \cref{fig:fit_results_bcudJ0_C01}, \cref{fig:fit_results_bcudJ1_C01}, and \cref{fig:fit_results_bbus_C00078} to \cref{fig:fit_results_bcudJ1_F006}. In \cref{fig:chiral_extrapolation}, we plot these results as a function of $m_\pi^2$.

\begin{table}[htb]
	\centering
	\begin{tabular}{lccccc} \hline \hline
		Ensemble & $ \bar{b}\bar{b}us $ && $ \bar{b}\bar{c}ud $, $ J=0 $ && $ \bar{b}\bar{c}ud $, $ J=1 $ \\
				 & $ \Delta E_0 $ [MeV] &\hspace{2ex}& $ \Delta E_0 $ [MeV] & \hspace{2ex}&$ \Delta E_0 $ [MeV] \\ \hline
		C00078 & $ -77(30) $ && $\:\, -   39(43) $ && $\:\, - 30(47) $  \\
		C005   & $ -76(22) $ && $\vp   104(47) $ && $\vpz 79(35) $ \\
		C01    & $ -83(24) $ && $\vpz   43(29) $ && $\vpz 40(31) $ \\
		F004   & $ -92(15) $ && $\vpz\vz 9(24) $ && $\vpz 21(40) $ \\
		F006   & $ -67(12) $ && $\vp   101(29) $ && $\vp 113(24) $ \\
		\hline \hline
	\end{tabular}
\caption{\label{tab:finalResults}Ground-state energies relative to the lowest meson-meson thresholds for the three investigated four-quark systems and for all five ensembles, i.e.\ $ \Delta E_0 = E_0 - E_{B} - E_{B_s^*} $ for $ \bar{b}\bar{b}us $, $ \Delta E_0 = E_0 - E_{B} - E_{D} $ for $ \bar{b}\bar{c}ud $ with $ J=0 $, and $ \Delta E_0 = E_0 - E_{B^*} - E_{D} $ for $ \bar{b}\bar{c}ud $ with $ J=1 $. }
\end{table}

% **********

\subsubsection{$ \bar{b}\bar{b}us $ with $ J^P=1^+ $}

For the $ \bar{b}\bar{b}us $ system we found ground-state energies around $70 \, \text{MeV}$ to $100 \, \text{MeV}$ below the $B B_s^\ast$ threshold. These are the energies in a finite periodic spatial volume of linear extent $N_s a \approx 2.7 \, \text{fm}$ for ensembles C005, C01, F004 and F006 and $N_s a \approx 5.3 \, \text{fm}$ for ensemble C00078. To extrapolate to infinite volume, we could, in principle, proceed as in our previous work \cite{Leskovec:2019ioa} on the $\bar b \bar b u d$ tetraquark with $I(J^P) = 0(1^+)$ and use L\"uscher's finite volume method \cite{Luscher:1990ux,Briceno:2017max}. For the $ \bar{b}\bar{b}us $ system this is, however, technically more complicated, because one has to take into account at least two scattering channels, $B B_s^\ast$ and $B^\ast B_s$, which have almost the same threshold energy. Moreover, the energy levels of the corresponding excitations are difficult to determine, as discussed in \cref{SEC562}. However, since the finite-volume ground-state energies are significantly below these thresholds, we expect only mild finite-volume corrections, much smaller than our current statistical errors. This expectation is supported by our infinite-volume extrapolations of $\bar b \bar b u d$ results in Ref.\ \cite{Leskovec:2019ioa}, where the finite-volume ground-state energies turned out to be essentially identical to their infinite-volume counterparts. Thus, we do not carry out an infinite-volume extrapolation in this work, but postpone such an analysis until we have improved lattice data available, in particular correlation functions with scattering operators at both the source and the sink.

\begin{figure}[!htb]
	\centering
	\includegraphics[width=0.49\linewidth, trim = 0 0 0 0 ,clip]{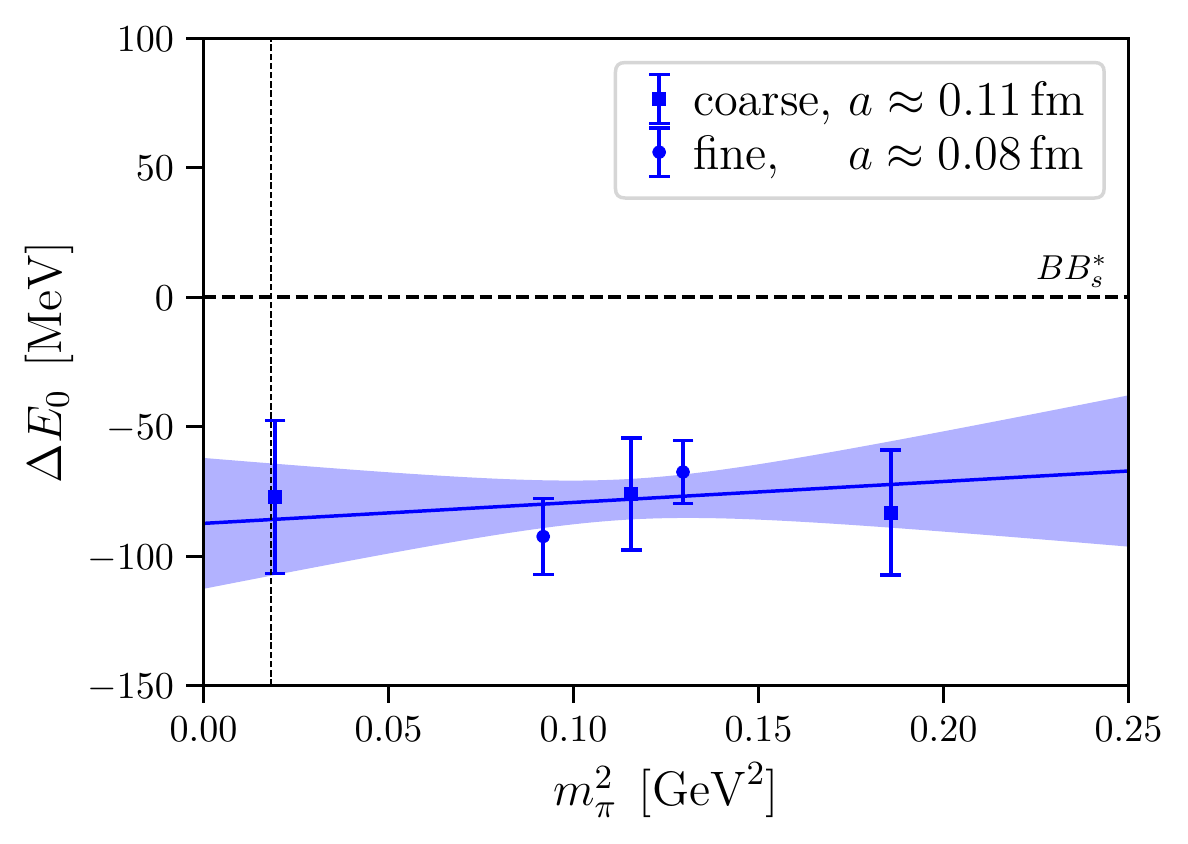} \\
	\includegraphics[width=0.49\linewidth, trim = 0 0 0 0 ,clip]{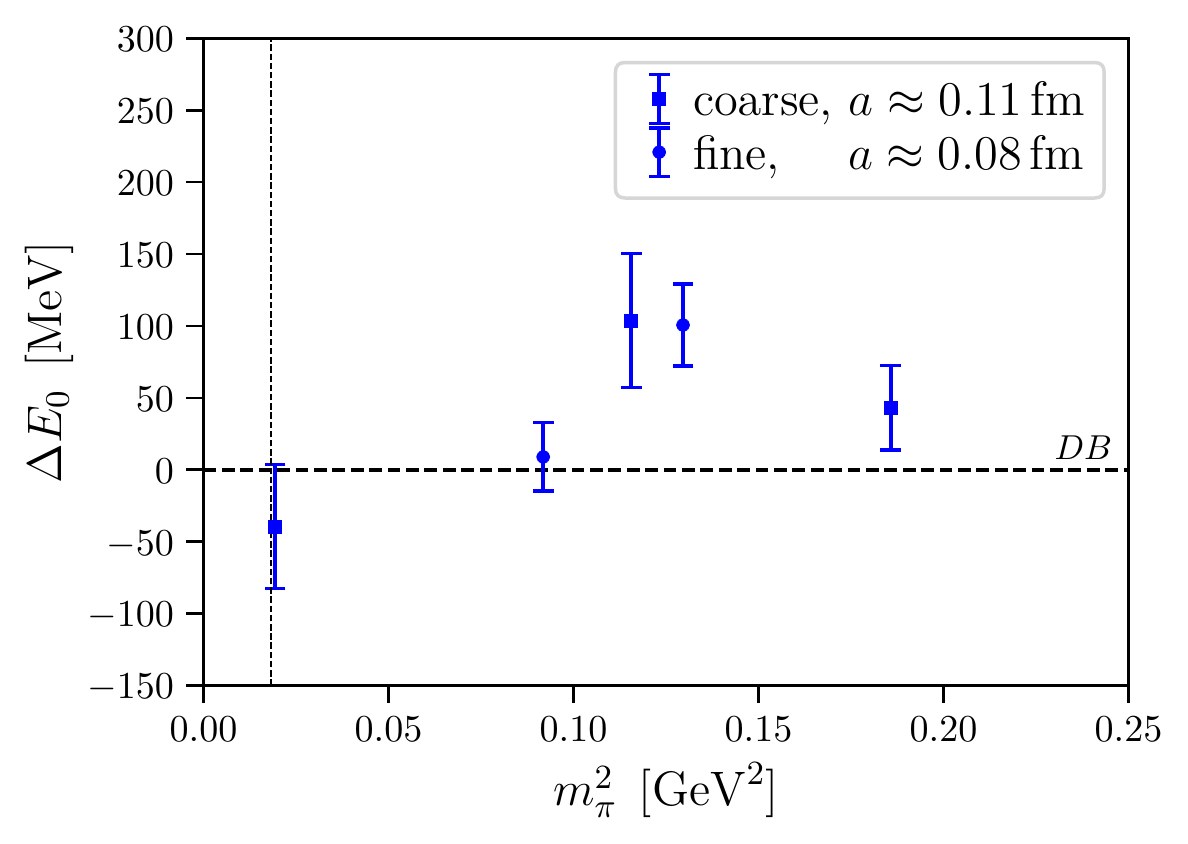}
	\includegraphics[width=0.49\linewidth, trim = 0 0 0 0 ,clip]{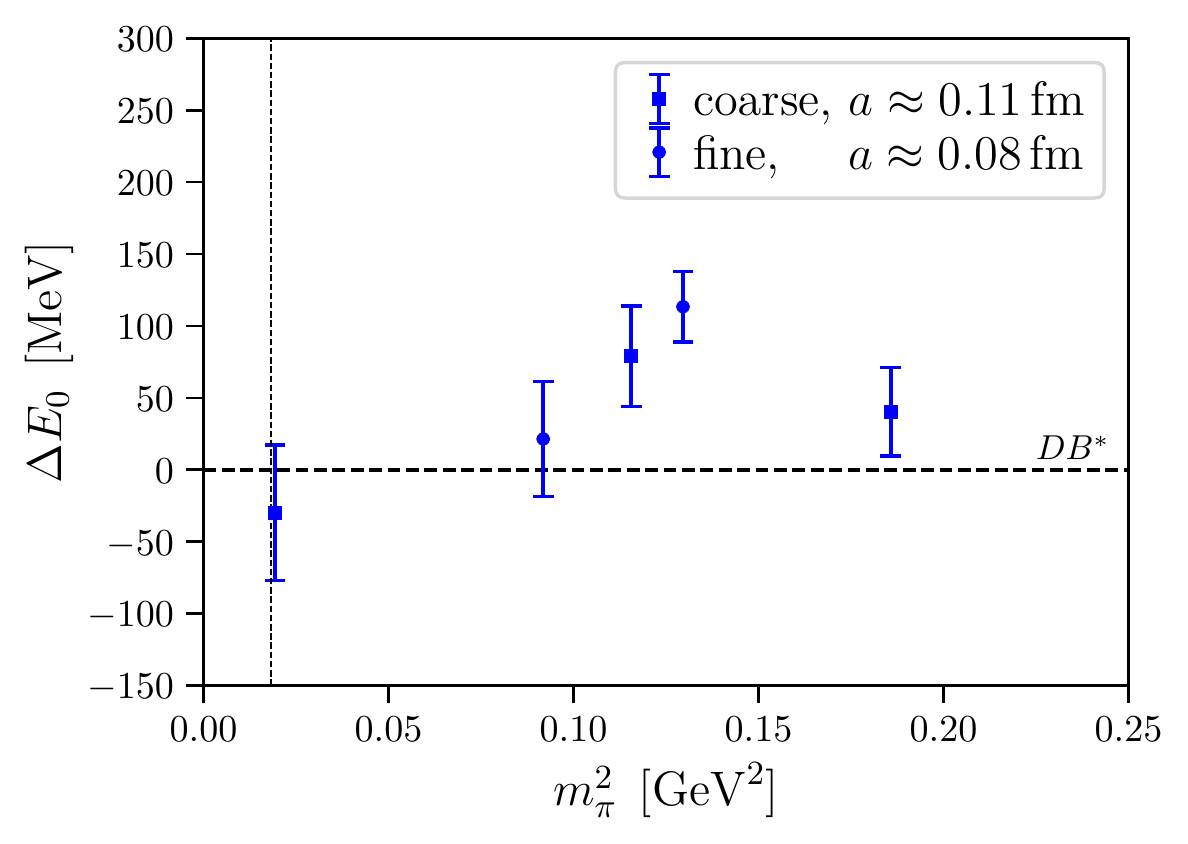}
	\caption{\label{fig:chiral_extrapolation}Ground-state energy as function of the squared pion mass for the $ \bar{b} \bar{b} u s $ system (top), the $ \bar{b} \bar{c} u d $ system with $ I(J^P) = 0(0^+) $ (bottom left) and the $ \bar{b} \bar{c} u d $ system with $ I(J^P) = 0(1^+) $ (bottom right). For $ \bar{b} \bar{b} u s $ we also show the fit and linear extrapolation to the physical point at $ m_{\pi,\textrm{phys}} = 135 \,\textrm{MeV} $ [see \cref{EQN_fit1} and \cref{EQN_fit2}]. Horizontal dashed lines indicate the lowest corresponding thresholds: the $B B_s^\ast$ threshold for $ \bar{b} \bar{b} u s $, the $B D$ threshold for $ \bar{b} \bar{c} u d $ with $ I(J^P) = 0(0^+) $, and the $B^\ast D$ threshold for $ \bar{b} \bar{c} u d $ with $ I(J^P) = 0(1^+) $.}
\end{figure}

Our five ensembles differ in the light-quark mass, corresponding to pion masses in the range $139 \, \text{MeV}$$\ltapprox m_\pi \ltapprox$ $431 \, \text{MeV}$, which allows us to perform an extrapolation of $\Delta E_0 = E_0 - E_{B} - E_{B_s^*}$ to the physical point (note that one of our ensembles, C00078, has a light quark mass that is almost physical). Since the observed dependence on the light-quark mass is mild (in fact, consistent with no dependence), a fit that is linear in $m_{u;d}$ and hence quadratic in $m_\pi^2$ is sufficient. We performed a $ \chi^2 $-minimizing fit using the ansatz
\begin{equation}
\label{EQN_fit1} \Delta E_0(m_\pi) = \Delta E_0(m_{\pi,\textrm{phys}}) + c \Big(m_{\pi}^2 - m_{\pi,\textrm{phys}}^2\Big) ,
\end{equation} 
where $\Delta E_0(m_{\pi,\textrm{phys}})$ and $c$ are fit parameters and $ m_{\pi,\textrm{phys}} = 135 \,\textrm{MeV} $. The resulting values for these parameters are
\begin{equation}
\label{EQN_fit2} \Delta E_0(m_{\pi,\textrm{phys}}) = (-86 \pm 22) \, \text{MeV} \quad , \quad c = (0.8 \pm 2.1) \times 10^{-4} / \text{MeV}^2
\end{equation} 
with $ \chi^2/\text{d.o.f.} = 0.81$, indicating consistency of the lattice data with our linear ansatz. The data points and the fit are shown in the upper plot of \cref{fig:chiral_extrapolation}.

There are also systematic errors due to the finite lattice spacing and the NRQCD action. We expect these errors to be of the same order as for the related $ \bar{b}\bar{b}ud $ system with $I(J^P) = 0(1^+)$. We have discussed these errors in detail in Section~VII our previous work \cite{Leskovec:2019ioa} and estimated them to be not larger than $10 \, \text{MeV}$. Thus, our final results for the $ \bar{b}\bar{b}us $ tetraquark binding
energy and mass are
\begin{equation}
\Delta E_0(m_{\pi,\textrm{phys}}) = (-86 \pm 22 \pm 10) \, \text{MeV} \quad , \quad m_{\bar{b}\bar{b}us \text{ tetraquark}}(m_{\pi,\textrm{phys}}) = (10609 \pm 22 \pm 10) \, \text{MeV},
\end{equation}
where $m_{\bar{b}\bar{b}us \text{ tetraquark}}$ is obtained by adding the experimental results of the $B$ and $B_s^\ast$ masses \cite{ParticleDataGroup:2020ssz} to $\Delta E_0$.
% mB ~ 5279.5, mBs* ~ 5415.4 --> threshhold ~ 10695

% **********

\subsubsection{$ \bar{b}\bar{c}ud $ with $ I(J^P)=0(0^+) $ and $ I(J^P)=0(1^+) $}

For both $ \bar{b} \bar{c} u d $ systems, the finite-volume ground-state energies are compatible with the corresponding lowest meson-meson thresholds. Thus, there is no indication that strong-interaction-stable tetraquarks exist in these channels. However, because of the statistical uncertainties of order $20 \, \text{MeV} \ldots 50 \, \text{MeV}$ (see \cref{tab:finalResults}), we cannot exclude the existence of a shallow bound state with binding energy of only a few MeV below the respective threshold.

Since we are not in a position to quantify finite-volume corrections, which might be sizable in particular for states close to the threshold, we also refrain from extrapolating our lattice results to physical pion mass. To summarize our finite-volume results in a graphical way, we nevertheless plot them in \cref{fig:chiral_extrapolation} in the same style as their $ \bar b \bar b u s $ counterparts together with the relevant meson-meson thresholds.

% ********************
% ********************
% ********************
% ********************
% ********************

\section{\label{SEC596}Conclusions and outlook}

We investigated a $\bar b \bar b u s$ and two $\bar b \bar c u d$ four-quark systems using lattice QCD with dynamical domain-wall $u$, $d$, and $s$ quarks. The charm quarks were implemented using an anisotropic clover action with parameters tuned to remove heavy-quark discretization errors, while the $b$ quarks were discretized within the framework of NRQCD. Our work improves upon existing similar studies \cite{Francis:2016hui,Francis:2018jyb,Junnarkar:2018twb,Hudspith:2020tdf,Padmanath:2021qje} by including also non-local (scattering) interpolating operators.

\begin{figure}[htb]
	\centering
	\includegraphics[width=0.85\linewidth]{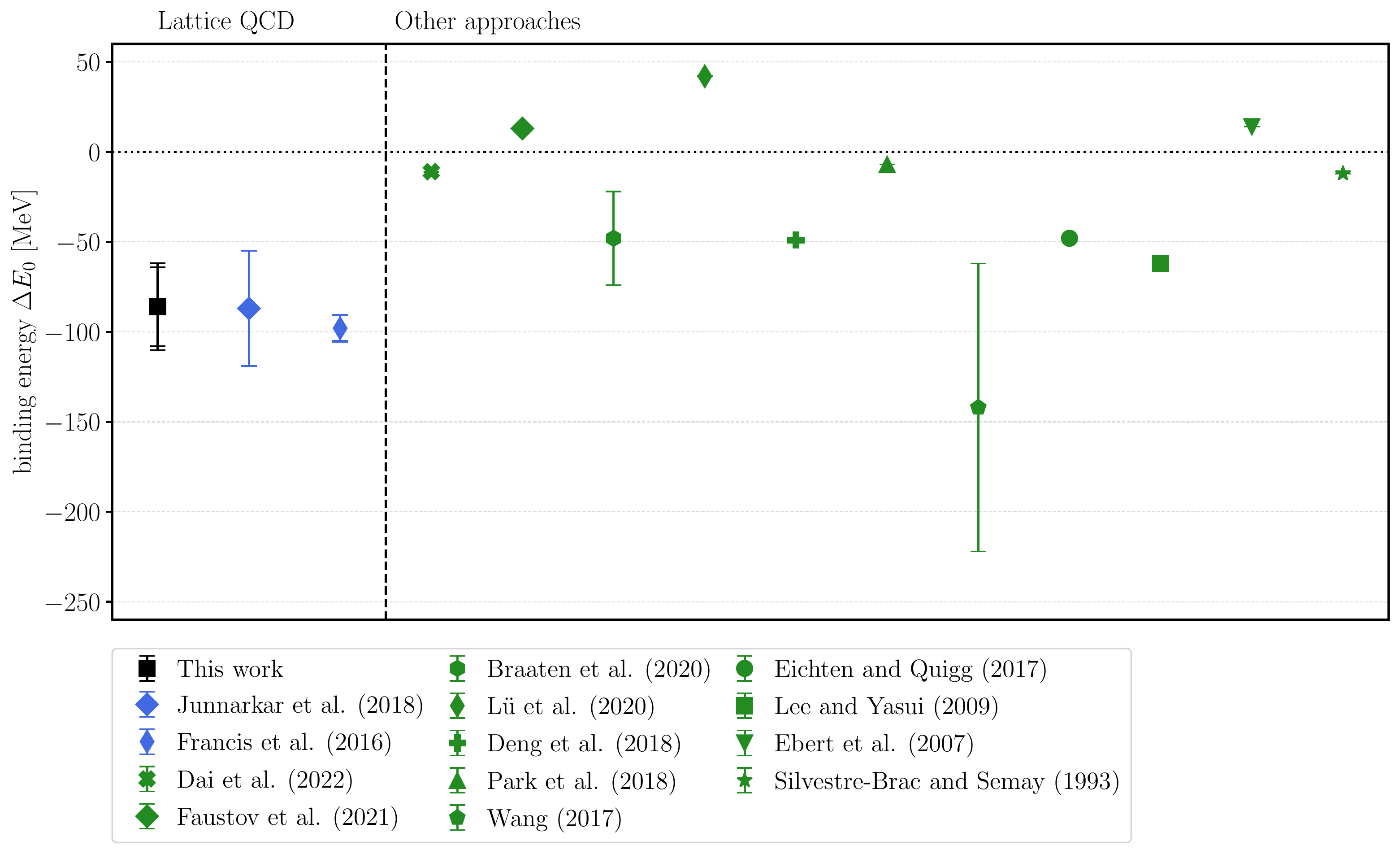}
	\caption{\label{FIG_summary}Comparison of results for the binding energy of the $\bar b \bar b u s$ tetraquark with $J^P = 1^+$ (black: this work, using lattice QCD; blue: previous works using lattice QCD \cite{Francis:2016hui,Junnarkar:2018twb}; green: other approaches (quark models, phenomenological considerations, sum rules) \cite{SilvestreBrac:1993ss,Ebert:2007rn,Lee:2009rt,Eichten:2017ffp,Wang:2017uld,Park:2018wjk,Deng:2018kly,Braaten:2020nwp,Lu:2020rog,Faustov:2021hjs,Dai:2022ulk}.}
\end{figure}

In the $\bar b \bar b u s$ sector with quantum numbers $J^P = 1^+$, we find clear evidence for a strong-interaction-stable tetraquark. The binding energy with respect to the $B B_s^\ast$ threshold is $(-86 \pm 22 \pm 10) \, \text{MeV}$, which is consistent with previous lattice QCD results from Refs.\ \cite{Francis:2016hui,Junnarkar:2018twb}. In \cref{FIG_summary} we summarize and compare these lattice QCD results with results obtained using different approaches, e.g.\ quark models, phenomenological considerations, or sum rules \cite{SilvestreBrac:1993ss,Ebert:2007rn,Lee:2009rt,Eichten:2017ffp,Wang:2017uld,Park:2018wjk,Deng:2018kly,Braaten:2020nwp,Lu:2020rog,Faustov:2021hjs,Dai:2022ulk}. As discussed in the introduction, there are strong discrepancies, even on a qualitative level, between these non-lattice results. Thus, it is important to have multiple independent first-principles lattice-QCD computations, and the agreement of the lattice results from different groups, as shown with the blue and black data points in \cref{FIG_summary}, increase the confidence in these results.

For the $\bar b \bar c u d$ systems with quantum numbers $I(J^P) = 0(0^+)$ and $I(J^P) = 0(1^+)$ the situation is less clear. We find finite-volume ground-state energies that are compatible with the lowest thresholds corresponding to $B D$ and $B^\ast D$, respectively. To decide whether there is a shallow bound state, more precise data and infinite-volume extrapolations will be needed. Results from previous lattice QCD studies \cite{Francis:2018jyb,Hudspith:2020tdf,Padmanath:2021qje} are mostly consistent with our results, but are also inconclusive. It is interesting to note that Ref.\ \cite{Padmanath:2021qje} reports a ground-state energy for $I(J^P) = 0(1^+)$ below the $B^\ast D$ threshold for a fine lattice spacing $a \approx 0.06 \, \text{fm}$, but not for the coarse lattice spacing $a \approx 0.12 \, \text{fm}$. The authors of Ref.\ \cite{Padmanath:2021qje} conclude that taking the continuum limit might be essential for the $\bar b \bar c u d$ system. We do not observe such a trend (see \cref{fig:chiral_extrapolation}), but it should be kept in mind that the types of lattice actions used here differ from Ref.\ \cite{Padmanath:2021qje}, except for the bottom quarks. As discussed in the introduction, also non-lattice studies do not clarify the possible existence of a strong-interaction-stable $\bar b \bar c u d$ tetraquark, since they exhibit strong discrepancies (Refs.\ \cite{Lee:2009rt,Chen:2013aba,Karliner:2017qjm,Sakai:2017avl,Agaev:2018khe,Deng:2018kly,Carames:2018tpe,Yang:2019itm,Tan:2020ldi} predict the existence of a stable tetraquark, while Refs.\ \cite{Ebert:2007rn,Eichten:2017ffp,Park:2018wjk,Braaten:2020nwp,Lu:2020rog} claim the opposite).

Our main goal for the future is to include scattering interpolating operators at both the sources and the sinks of our correlation matrices (rather than just the sinks as done here). We expect that this will allow us to determine the low-lying energy levels, in particular those associated with scattering states, more reliably and more precisely. We could then carry out infinite-volume extrapolations for the $\bar b \bar c u d$ systems using L\"uscher's method \cite{Luscher:1990ux} and possibly clarify the existence or non-existence of a strong-interaction-stable $\bar b \bar c u d$ tetraquark. Another interesting direction could be to explore heavy-heavy-light-light four-quark systems with other quantum numbers for which stable tetraquarks are not expected, but for which resonances could exist. A clear candidate is the $\bar b \bar b u d$ system with $I(J^P) = 0(1^-)$, where such a resonance around $15 \, \text{MeV}$ above the $B B$ threshold was predicted using static-static-light-light potentials computed with lattice QCD and the Born-Oppenheimer approximation \cite{Bicudo:2017szl}.

% ********************
% ********************
% ********************
% ********************
% ********************

\FloatBarrier

\section*{Acknowledgements}

We thank the RBC and UKQCD collaborations for providing the gauge-link ensembles. We thank Luka Leskovec for collaboration on earlier related work. We also acknowledge useful discussions with Ahmed Ali.

S.M.\ is supported by the U.S.\ Department of Energy, Office of Science, Office of High Energy Physics under Award Number D{E-S}{C0}009913.
M.P.\ and M.W.\ acknowledge support by the Deutsche Forschungsgemeinschaft (DFG, German Research Foundation) -- project number 457742095.
M.W.\ acknowledges support by the Heisenberg Programme of the Deutsche Forschungsgemeinschaft (DFG, German Research Foundation) -- project number 399217702.

This research used resources of the National Energy Research Scientific Computing Center (NERSC), a U.S.\ Department of Energy Office of Science User Facility operated under Contract No.\ DE-AC02-05CH11231. This work also used resources at the Texas Advanced Computing Center that are part of the Extreme Science and Engineering Discovery Environment (XSEDE), which is supported by National Science Foundation grant number ACI-1548562.
Calculations on the GOETHE-HLR and on the FUCHS-CSC  high-performance computers of the Frankfurt University were conducted for this research. We would like to thank HPC-Hessen, funded by the State Ministry of Higher Education, Research and the Arts, for programming advice.

% ********************
% ********************
% ********************
% ********************
% ********************

\appendix

\FloatBarrier

\section{\label{app:energytablesandfigures}Summary plots of multi-exponential fits to determine energy levels for ensembles C00078, C005, F004 and F006}

\begin{figure}[!htb]
	\centering
	\includegraphics[width=0.8\linewidth, trim = 0 0 0 0 ,clip]{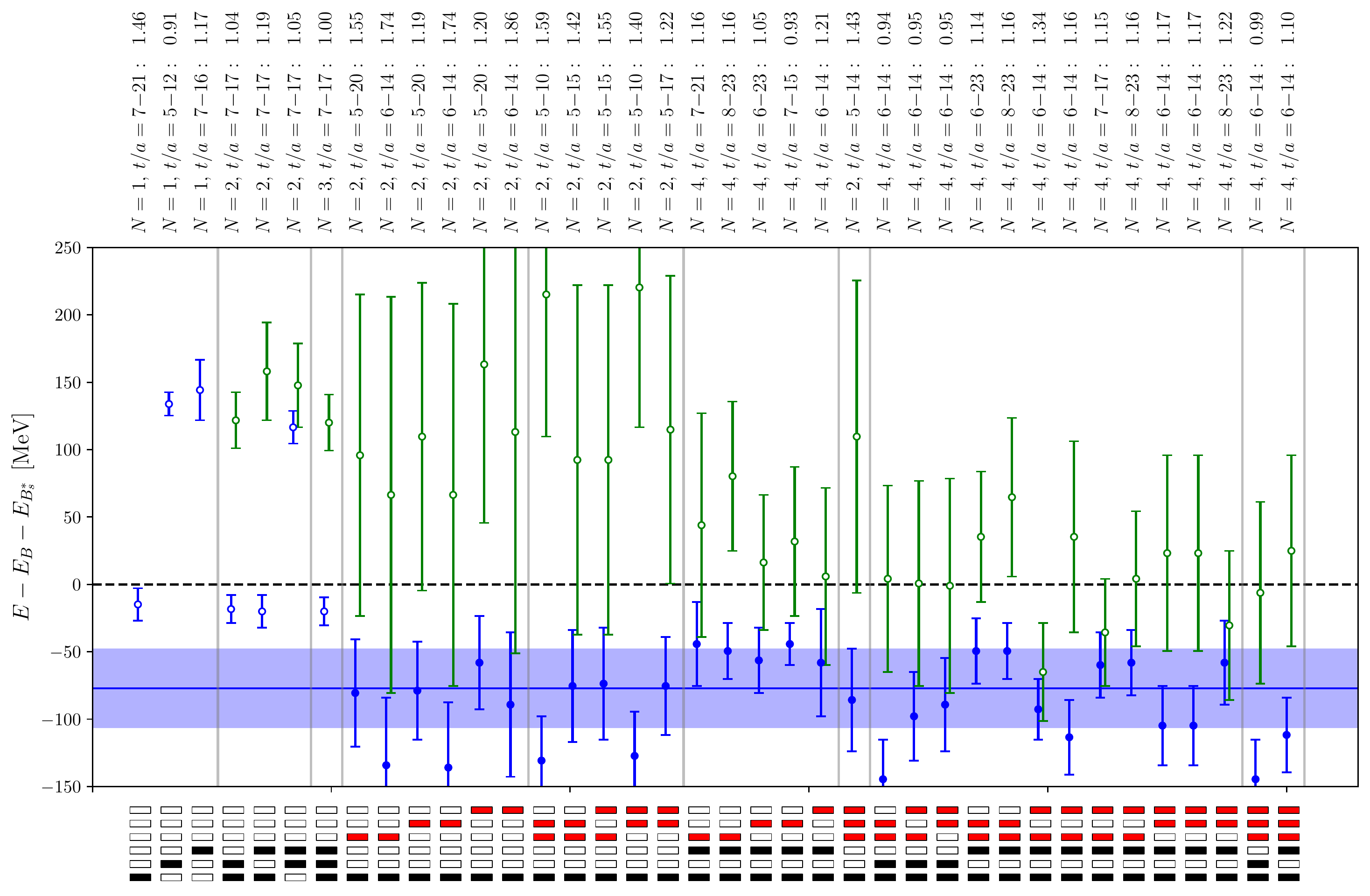} \vspace{-1ex}
	\caption{\label{fig:fit_results_bbus_C00078} Fit results for $E_0$ (blue) and $E_1$ (green) for the $ \bar{b}\bar{b}us $ system relative to the $ B B_s^\ast $ threshold for ensemble C00078.}
\end{figure}

\begin{figure}[!htb]
	\centering
	\includegraphics[width=0.9\linewidth, trim = 0 0 0 0 ,clip]{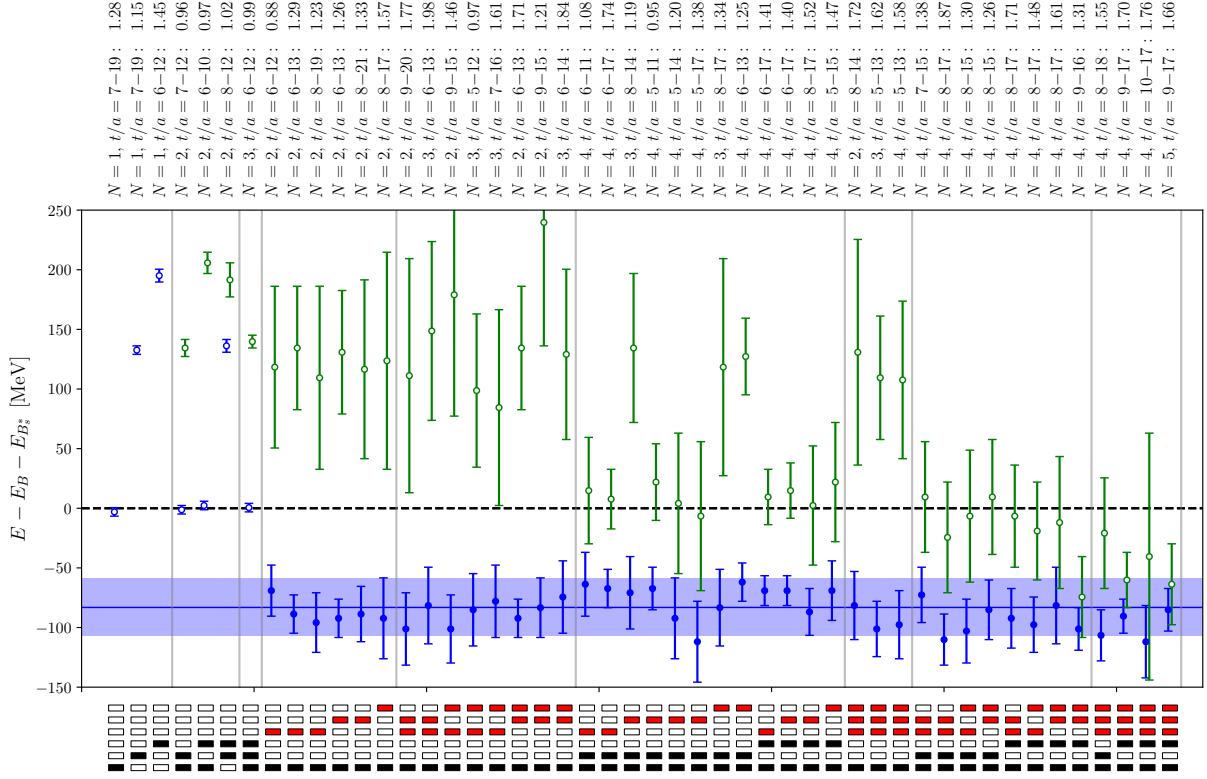} \vspace{-1ex}
	\caption{\label{fig:fit_results_bbus_C005} Fit results for $E_0$ (blue) and $E_1$ (green) for the $ \bar{b}\bar{b}us $ system relative to the $ B B_s^\ast $ threshold for ensemble C005.}
\end{figure}

\begin{figure}[!htb]
	\centering
	\includegraphics[width=0.88\linewidth, trim = 0 0 0 0 ,clip]{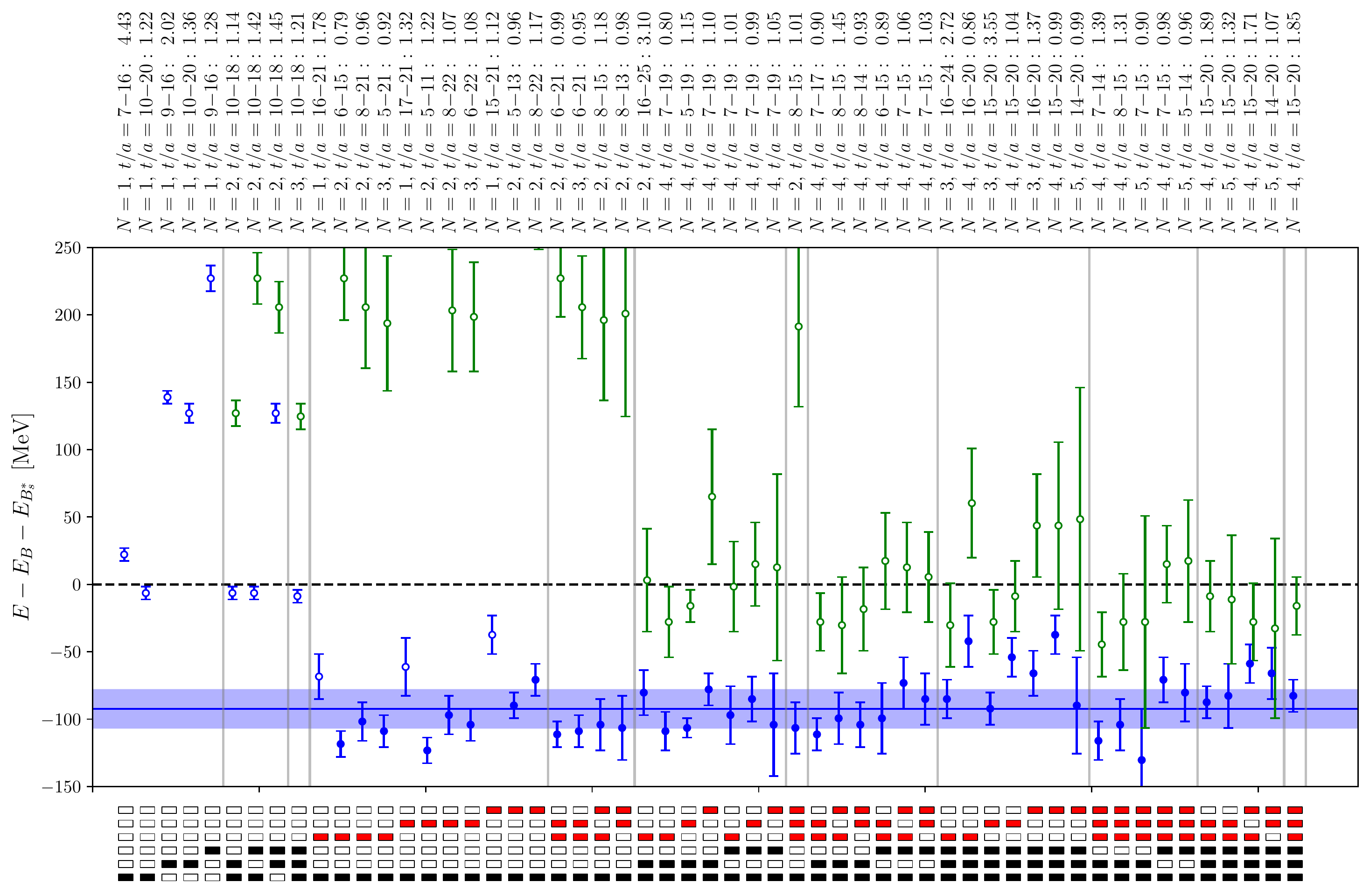} \vspace{-1ex}
	\caption{\label{fig:fit_results_bbus_F004} Fit results for $E_0$ (blue) and $E_1$ (green) for the $ \bar{b}\bar{b}us $ system relative to the $ B B_s^\ast $ threshold for ensemble F004.}
\end{figure}

\begin{figure}[!htb]
	\centering
	\includegraphics[width=0.88\linewidth, trim = 0 0 0 0 ,clip]{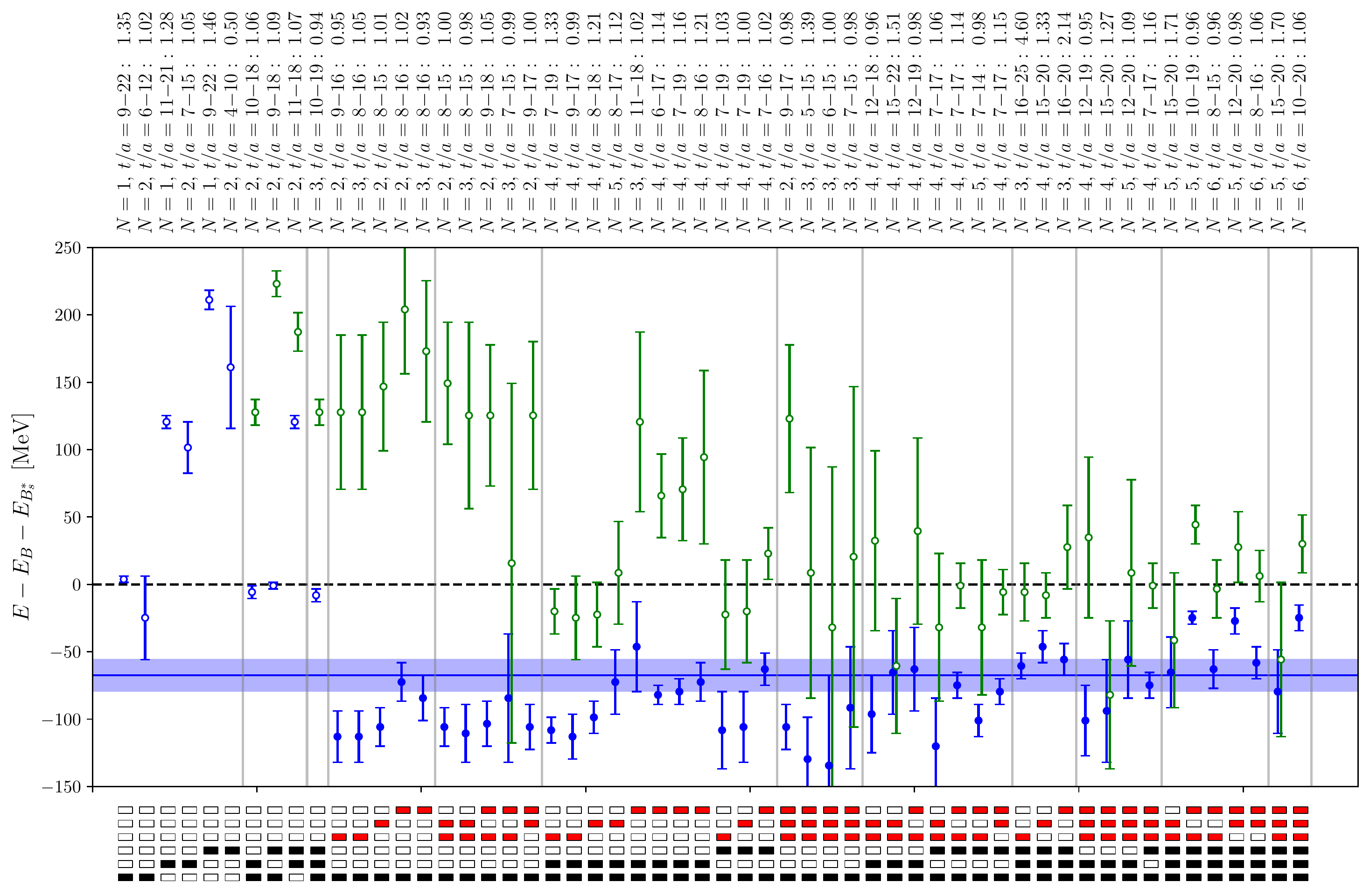} \vspace{-1ex}
	\caption{\label{fig:fit_results_bbus_F006} Fit results for $E_0$ (blue) and $E_1$ (green) for the $ \bar{b}\bar{b}us $ system relative to the $ B B_s^\ast $ threshold for ensemble F006.}
\end{figure}

\begin{figure}[!htb]
	\centering
	\includegraphics[width=0.55\linewidth, trim = 0 0 0 0 ,clip]{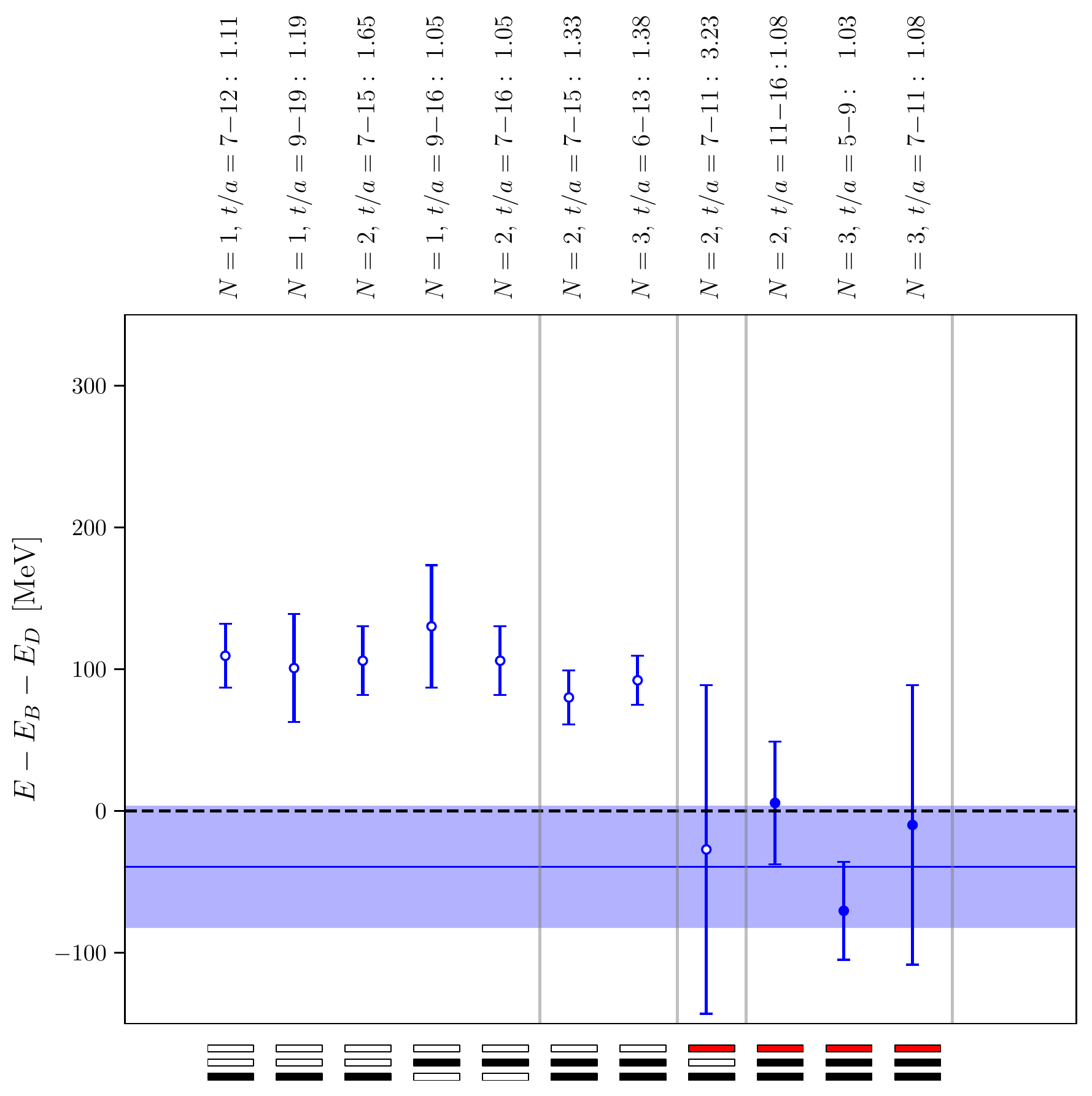}
	\caption{\label{fig:fit_results_bcudJ0_C00078} Fit results for $E_0$ for the $ \bar{b}\bar{c}ud $ system with $ I(J^P)=0(0^+) $ relative to the $ B D $ threshold for ensemble C00078.}
\end{figure}

\begin{figure}[!htb]
	\centering
	\includegraphics[width=0.55\linewidth, trim = 0 0 0 0 ,clip]{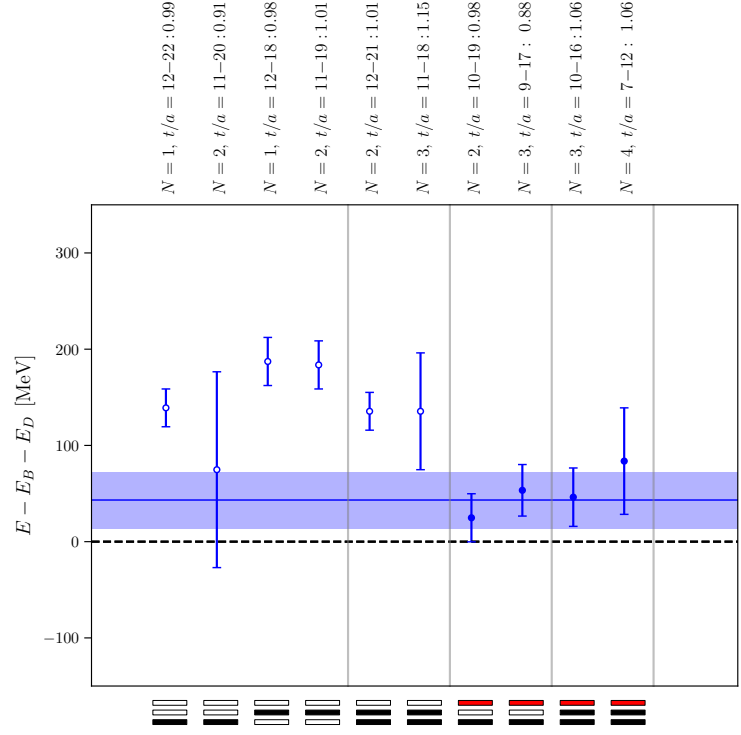}
	\caption{\label{fig:fit_results_bcudJ0_C005} Fit results for $E_0$ for the $ \bar{b}\bar{c}ud $ system with $ I(J^P)=0(0^+) $ relative to the $ B D $ threshold for ensemble C005.}
\end{figure}

\begin{figure}[!htb]
	\centering
	\includegraphics[width=0.55\linewidth, trim = 0 0 0 0 ,clip]{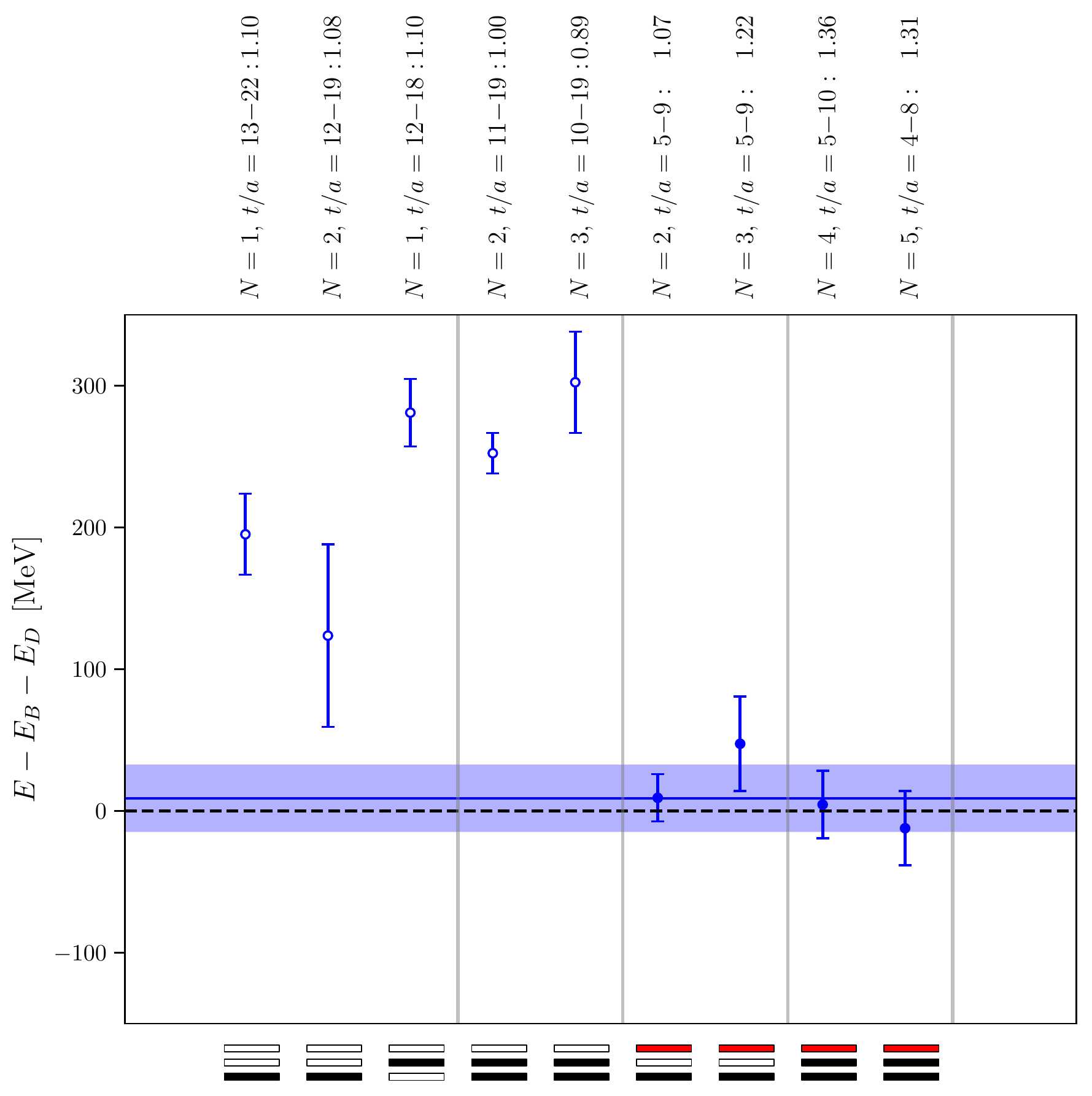}
	\caption{\label{fig:fit_results_bcudJ0_F004} Fit results for $E_0$ for the $ \bar{b}\bar{c}ud $ system with $ I(J^P)=0(0^+) $ relative to the $ B D $ threshold for ensemble F004.}
\end{figure}

\begin{figure}[!htb]
	\centering
	\includegraphics[width=0.55\linewidth, trim = 0 0 0 0 ,clip]{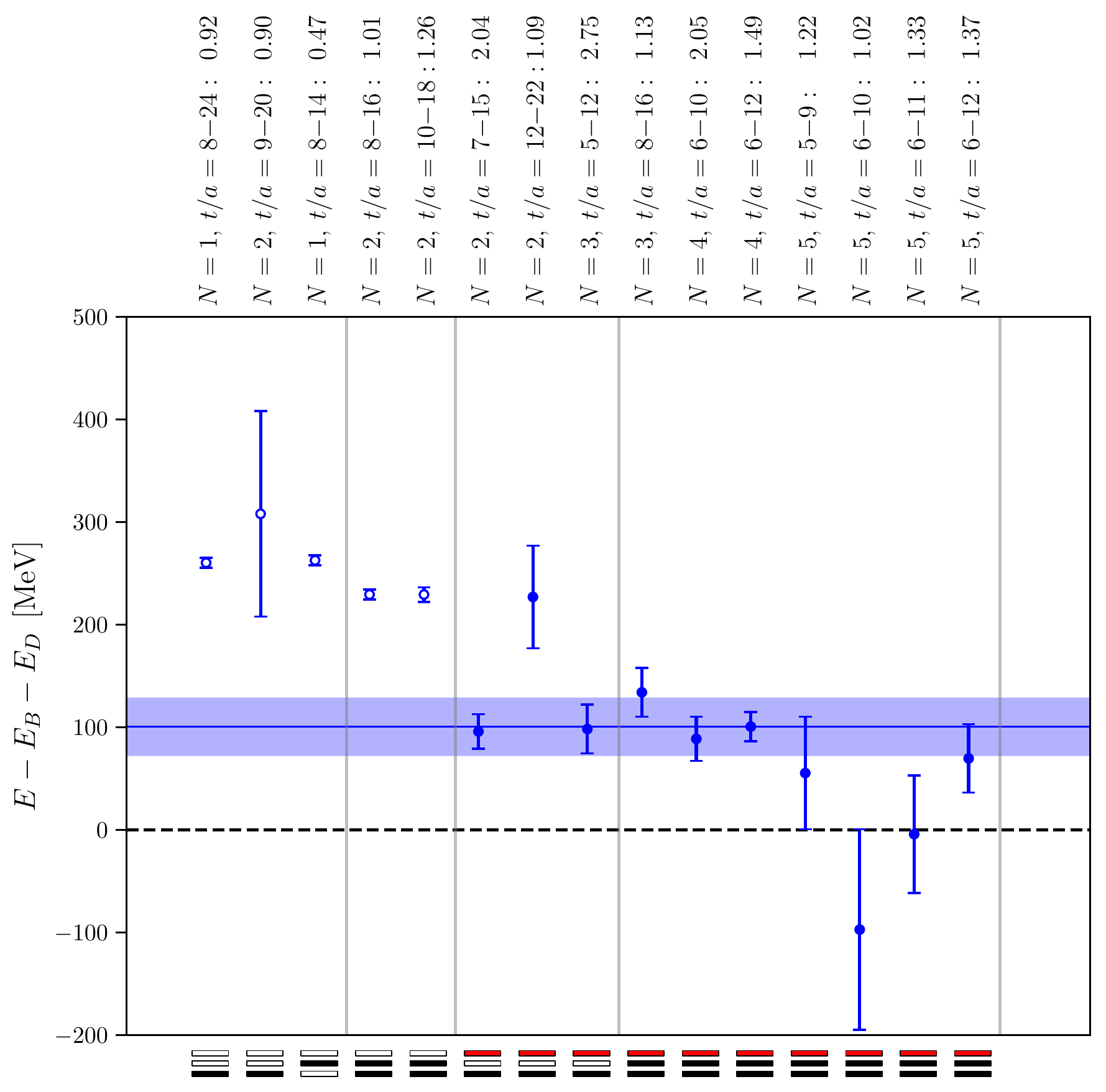}
	\caption{\label{fig:fit_results_bcudJ0_F006} Fit results for $E_0$ for the $ \bar{b}\bar{c}ud $ system with $ I(J^P)=0(0^+) $ relative to the $ B D $ threshold for ensemble F006.}
\end{figure}

\begin{figure}[!htb]
	\centering
	\includegraphics[width=0.65\linewidth, trim = 0 0 0 0 ,clip]{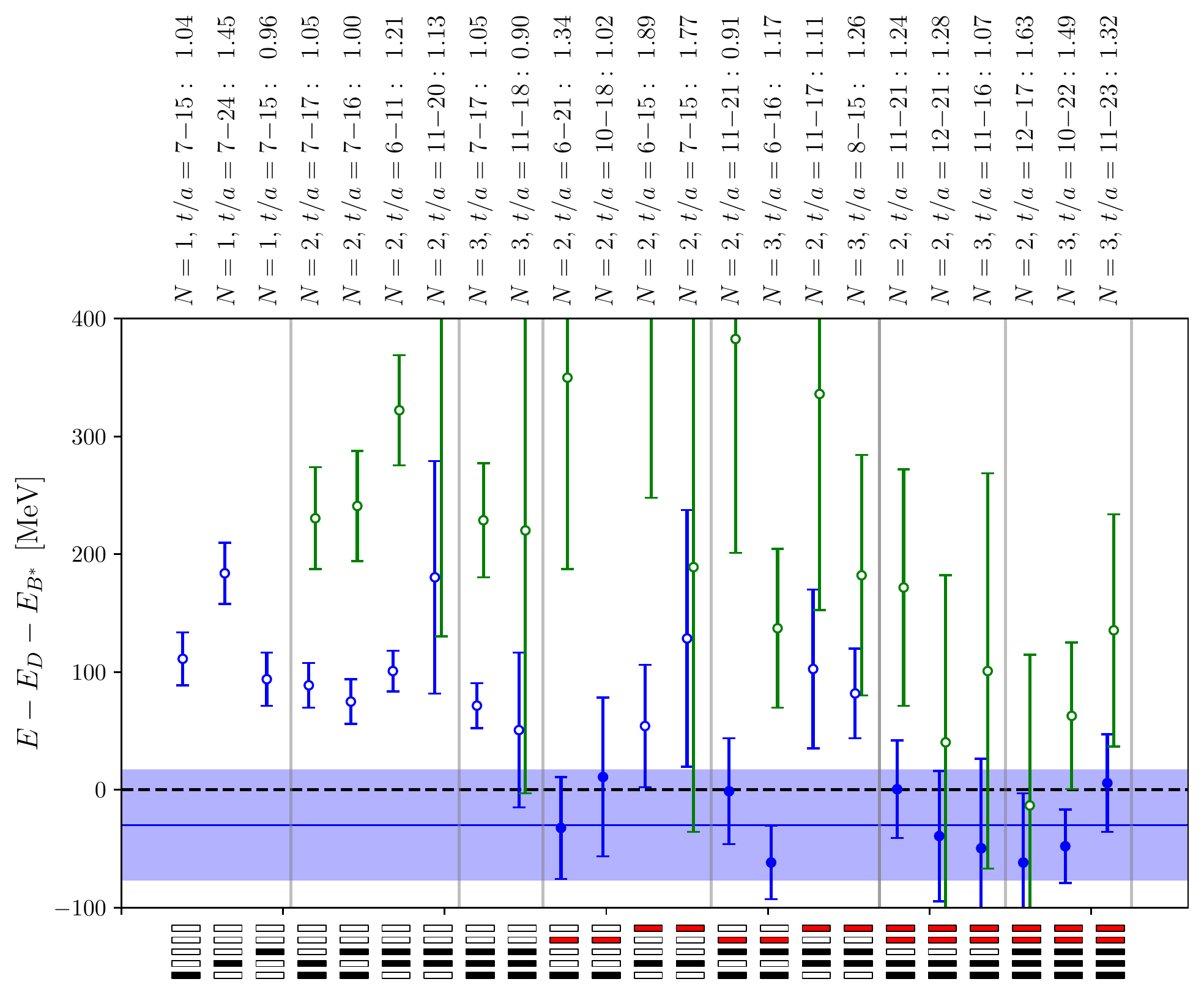}
	\caption{\label{fig:fit_results_bcudJ1_C00078} Fit results for $E_0$ (blue) and $E_1$ (green) for the $ \bar{b}\bar{c}ud $ system with $ I(J^P)=0(1^+) $ relative to the $ B^\ast D $ threshold for ensemble C00078.}
\end{figure}

\begin{figure}[!htb]
	\centering
	\includegraphics[width=0.65\linewidth, trim = 0 0 0 0 ,clip]{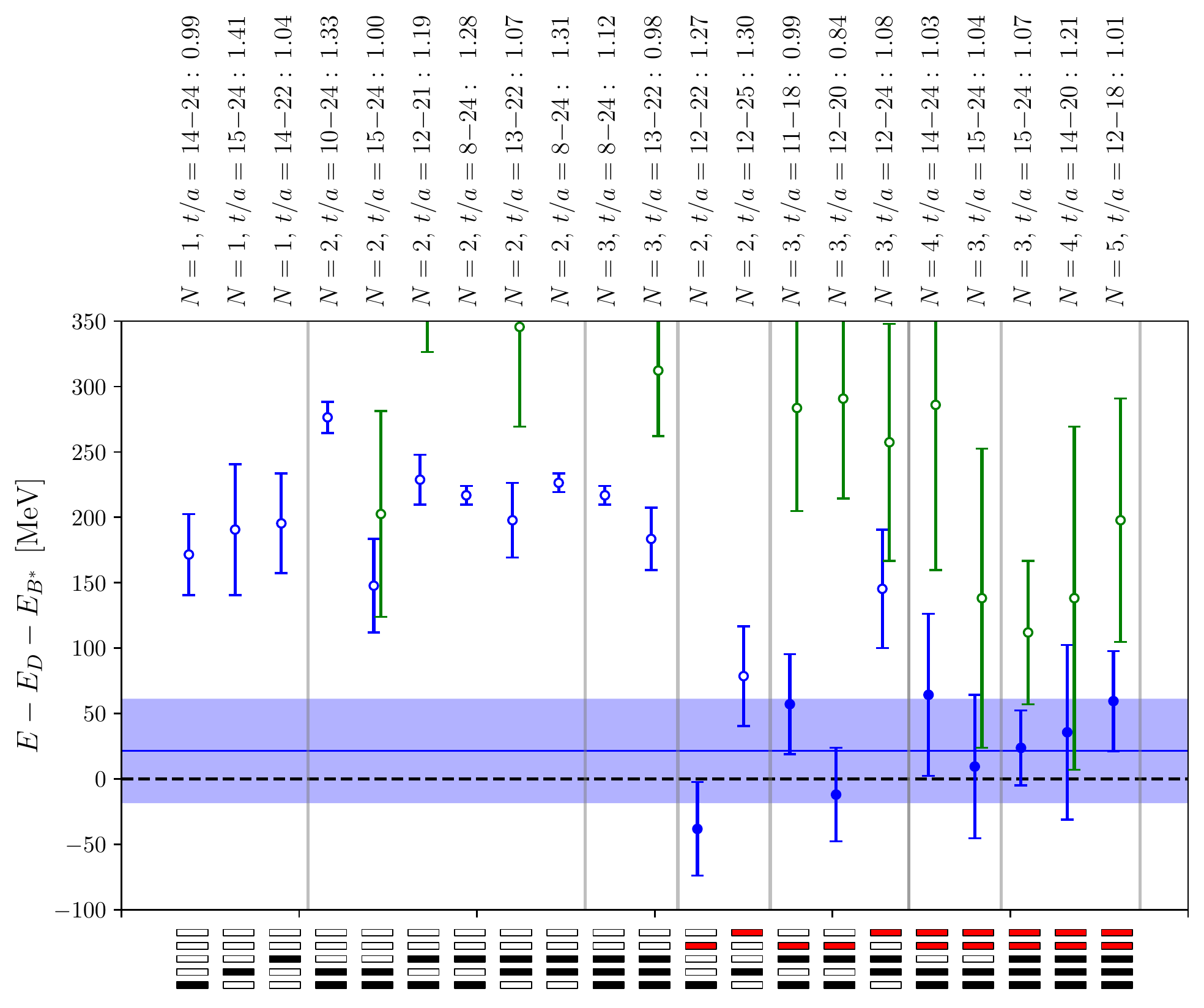}
	\caption{\label{fig:fit_results_bcudJ1_F004} Fit results for $E_0$ (blue) and $E_1$ (green) for the $ \bar{b}\bar{c}ud $ system with $ I(J^P)=0(1^+) $ relative to the $ B^\ast D $ threshold for ensemble F004.}
\end{figure}

\begin{figure}[!htb]
	\centering
	\includegraphics[width=0.65\linewidth, trim = 0 0 0 0 ,clip]{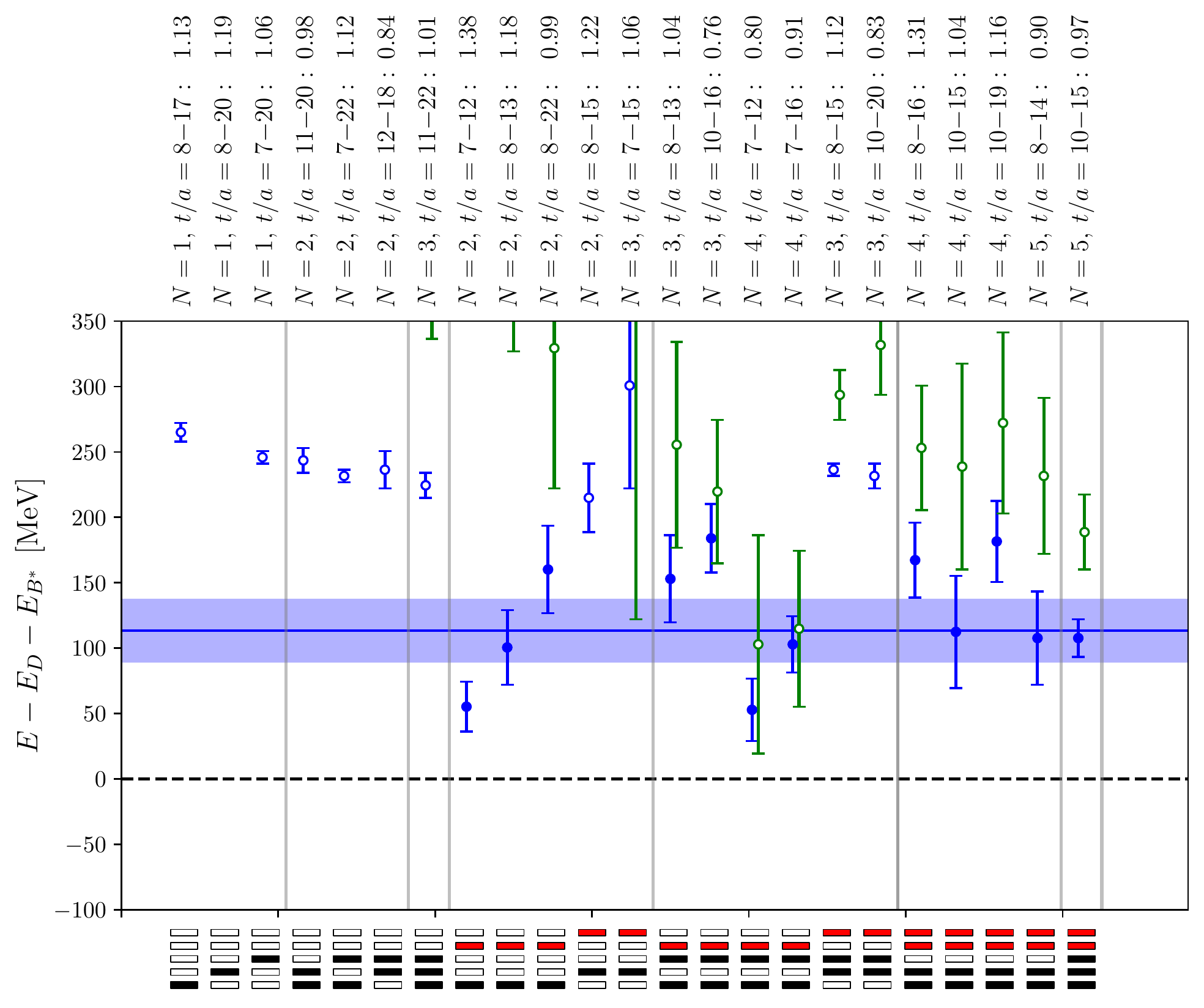}
	\caption{\label{fig:fit_results_bcudJ1_F006} Fit results for $E_0$ (blue) and $E_1$ (green) for the $ \bar{b}\bar{c}ud $ system with $ I(J^P)=0(1^+) $ relative to the $ B^\ast D $ threshold for ensemble F006.}
\end{figure}

In this appendix we show the results of multi-exponential fits to determine $E_0$ and $E_1$ for the ensembles C00078, C005, F004, and F006:
\begin{itemize}
\item $ \bar{b}\bar{b}us $ with $ J^P=1^+ $: \cref{fig:fit_results_bbus_C00078} to \cref{fig:fit_results_bbus_F006}.

\item $ \bar{b}\bar{c}ud $ with $ I(J^P)=0(0^+) $: \cref{fig:fit_results_bcudJ0_C00078} to \cref{fig:fit_results_bcudJ0_F006}.

\item $ \bar{b}\bar{c}ud $ with $ I(J^P)=0(1^+) $: \cref{fig:fit_results_bcudJ1_C00078} to \cref{fig:fit_results_bcudJ1_F006}.
\end{itemize}
The style of these figures is identical to \cref{fig:fit_results_bbus_C01}, \cref{fig:fit_results_bcudJ0_C01} and \cref{fig:fit_results_bcudJ1_C01}, respectively, where the same quantities are shown for ensemble C01, and which are discussed in detail in \cref{sec:results}.

% *****

% *****

% *****

\begin{figure}[!htb]
	\centering
	\includegraphics[width=0.65\linewidth, trim = 0 0 0 0 ,clip]{figs/fitresults_C01_bcud_J1_preliminary_v2_1state.pdf}
	\caption{\label{fig:fit_results_bcudJ1_C005} Fit results for $E_0$ (blue) and $E_1$ (green) for the $ \bar{b}\bar{c}ud $ system with $ I(J^P)=0(1^+) $ relative to the $ B^\ast D $ threshold for ensemble C005.}
\end{figure}

% ********************
% ********************
% ********************

\FloatBarrier

\section{\label{SEC607}Method to estimate the ground state energy from several multi-exponential fits}

To obtain a final estimate of the ground-state energy and its uncertainty from several different selected multi-exponential fits, we follow the approach of the FLAG collaboration, discussed e.g.~in Section~2.3.1 of their 2019 review \cite{FlavourLatticeAveragingGroup:2019iem}. The starting point is the result of each fit given as
\begin{equation}
\label{EQN755} E_0^{(j)} \pm \Delta E_0^{(j)} ,
\end{equation}
where $j$ is the index of the fit, $E_0^{(j)}$ the mean value and $\Delta E_0^{(j)}$ the statistical error.

We estimate the ground state energy by a weighted average,
\begin{equation}
\bar{E}_0 = \sum_j \omega^{(j)} E_0^{(j)} .
\end{equation}
The weights are given by
\begin{equation}
\omega^{(j)} = \frac{1 / (\Delta E_0^{(j)})^2}{\sum_j 1/(S^{(j)} \Delta E_0^{(j)})^2} ,
\end{equation}
where $\sigma^{(j)} = S^{(j)} \Delta E_0^{(j)}$ with $S^{(j)} = \text{max}(1, (\chi_j^2/\text{d.o.f.})^{(j)})^{1/2})$. Thus, the estimate of $\bar{E}_0$ is equivalent to the result of a weighted, uncorrelated, $ \chi^2 $-minimizing fit of a constant to the results (\ref{EQN755}), where fits of bad quality, i.e.\ with $(\chi^2/\text{d.o.f.})^{(j)} > 1$, are additionally suppressed by $S^{(j)}$.

The selected multi-exponential fits are based on the same gauge link configurations and the same two-point functions and are, thus, correlated. The multi-exponential matrix fits are computationally demanding and a resampling procedure needed to quantify the correlations was not feasible. We there therefore conservatively assume the correlations to be maximal. The uncertainty of the ground state energy is then
\begin{equation}
\Delta \bar{E}_0 = \bigg(\sum_{j,k} \omega^{(j)} \omega^{(j)} \sigma^{(j)} \sigma^{(k)}\bigg)^{1/2} .
\end{equation}
The results $\bar{E}_0 \pm \Delta \bar{E}_0$ are shown as blue horizontal lines and light blue bands in \cref{fig:fit_results_bbus_C01}, \cref{fig:fit_results_bcudJ0_C01}, \cref{fig:fit_results_bcudJ1_C01}, and \cref{fig:fit_results_bbus_C00078} to \cref{fig:fit_results_bcudJ1_F006}.

% ********************

\providecommand{\href}[2]{#2}\begingroup\raggedright\endgroup

% ********************

\end{document}